\documentclass[twocolumn]{aastex631}

\usepackage{savesym}
\savesymbol{tablenum}
\usepackage{siunitx}
\restoresymbol{SIX}{tablenum}

\usepackage{wasysym}

\DeclareSIUnit\au{AU}
\DeclareSIUnit\msun{M_\odot}
\DeclareSIUnit\mearth{M_\oplus}
\DeclareSIUnit\year{yr}

\defcitealias{2019ApJCambioni}{C19}
\defcitealias{2020ApJEmsenhuberA}{E20}

\defcitealias{2019ApJEmsenhuberA}{Paper I}

\def\c19{\citetalias{2019ApJCambioni}}
\def\e20{\citetalias{2020ApJEmsenhuberA}}
\def\paperone{\citetalias{2019ApJEmsenhuberA}}

\received{2021 April 20}
\revised{2021 July 13}
\accepted{2021 July 20}
\published{2021 September 23}
\submitjournal{PSJ}

\shorttitle{Collision Chains among the Terrestrial Planets. II.}
\shortauthors{Emsenhuber et al.}

\begin{document}

\title{Collision Chains among the Terrestrial Planets. II. An Asymmetry between Earth and Venus}

\correspondingauthor{Alexandre Emsenhuber}
\email{emsenhuber@usm.lmu.de}

\author[0000-0002-8811-1914]{Alexandre Emsenhuber}
\affiliation{Lunar and Planetary Laboratory, University of Arizona, 1629 E. University Blvd., Tucson, AZ 85721, USA}
\affiliation{Space Research and Planetary Science, University of Bern, Gesellschaftsstrasse 6, 3012 Bern, Switzerland}
\affiliation{Universitäts-Sternwarte München, Ludwig-Maximilians-Universität München, Scheinerstraße 1, 81679 München, Germany}

\author[0000-0003-1002-2038]{Erik Asphaug}
\affiliation{Lunar and Planetary Laboratory, University of Arizona, 1629 E. University Blvd., Tucson, AZ 85721, USA}

\author[0000-0001-6294-4523]{Saverio Cambioni}
\affiliation{Lunar and Planetary Laboratory, University of Arizona, 1629 E. University Blvd., Tucson, AZ 85721, USA}

\author[0000-0002-9767-4153]{Travis S. J. Gabriel}
\affiliation{School of Earth and Space Exploration, Arizona State University, 781 E. Terrace Mall, Tempe, AZ 85287, USA}

\author[0000-0001-6294-4523]{Stephen R. Schwartz}
\affiliation{Lunar and Planetary Laboratory, University of Arizona, 1629 E. University Blvd., Tucson, AZ 85721, USA}

\begin{abstract}
During the late stage of terrestrial planet formation, hit-and-run collisions are about as common as accretionary mergers, for expected velocities and angles of giant impacts.
Average hit-and-runs leave two major remnants plus debris: the target and impactor, somewhat modified through erosion, escaping at lower relative velocity.
Here we continue our study of the dynamical effects of such collisions. We compare the dynamical fates of intact runners that start from hit-and-runs with proto-Venus at \SI{0.7}{au} and proto-Earth at \SI{1.0}{au}. We follow the orbital evolutions of the runners, including the other terrestrial planets, Jupiter, and Saturn, in an \textit{N}-body code. We find that the accretion of these runners can take $\gtrsim$\SI{10}{\mega\year} (depending on the egress velocity of the first collision) and can involve successive collisions with the original target planet or with other planets. We treat successive collisions that the runner experiences using surrogate models from machine learning, as in previous work, and evolve subsequent hit-and-runs in a similar fashion. We identify asymmetries in the capture, loss, and interchange of runners in the growth of Venus and Earth. Hit-and-run is a more probable outcome at proto-Venus, being smaller and faster orbiting than proto-Earth. But Venus acts as a sink, eventually accreting most of its runners, assuming typical events, whereas proto-Earth loses about half, many of those continuing to Venus.
This leads to a disparity in the style of late-stage accretion that could have led to significant differences in geology, composition, and satellite formation at Earth and Venus.
\end{abstract}


\section{Introduction}
\label{sec:intro}

Earth and Venus are referred to as ``sister planets,'' as they have a similar mass (Venus being only \SI{15}{\percent} less massive) and bulk density. Numerical simulations of the formation of the solar system's terrestrial planets reproduce the formation of Earth and Venus analogs under a wide variety of initial conditions \citep{2001IcarusChambers,2006IcarusOBrien,2006IcarusRaymond,2009RaymondIcarus,2009ApJHansen,2014EPSLFischerCiesla}. Yet Venus somehow ended up in a completely different dynamical state, rotating retrograde compared to the other planets and with a rotation period of 243 days \citep{2019IcarusCampbell}, with no known satellites \citep{2009IcarusSheppard}. The Earth-Moon system has more than \num{1000} times the angular momentum per unit mass than Venus.

Various explanations have been proposed for these differences, such as a despinning by tidal torques due to the planet's interior and its atmosphere \citep{2001NatureCorreiaLaskar,2017AAAuclairDesrotour}, or that it did not undergo giant impacts \citep{2017EPSLJacobson}. The slow rotation of Venus and its lack of an internal magnetic dynamo may be consistent with the absence of giant impacts \citep{2017EPSLJacobson}, or could indicate only head-on giant impacts with low angular momentum. However, giant impacts are predicted to dominate the late stage of planet formation \citep{1985ScienceWetherill,2002ApJKokubo}. The early accretion of planetesimals alone does not provide much net spin \citep{1990IcarusIdaNakazawa,1991IcarusLissauerSafronov,1993IcarusDonesTremaine}, while giant impacts can strongly augment angular momentum \citep[e.g.][]{1999IcarusAgnor}. There are also other formation models where there may have only been one giant impact, namely, the one responsible for the formation of the Moon \citep{2021SciAdvJohansen}.

Giant impacts are collisions between similar-sized planetary embryos occurring at velocities $v_\mathrm{coll}$ that are comparable to the mutual escape velocity
\begin{equation}
v_\mathrm{esc} = \sqrt{2G(m_\mathrm{tar}+m_\mathrm{imp})/(r_\mathrm{tar}+r_\mathrm{imp})}.
\end{equation}
Here $m_\mathrm{tar}\geq m_\mathrm{imp}$ are the masses of the target and the impactor, $r_\mathrm{tar}$ and $r_\mathrm{imp}$ the corresponding radii, and $G$ is the gravitation constant. Late-stage collisions therefore occur at a few to $\gtrsim$\SI{10}{\kilo\meter\per\second} for nominal terrestrial embryos, which is comparable to the sound speed in ices, oxides, and metals \citep{2015AstIVAsphaug}, resulting in global shocks that can generate magma oceans \citep{1993JGRTonksMelosh} and trigger differentiation \citep[see][]{1979JGRKaula,2015IcarusRubie}. Nonetheless, simulations show that the gross outcomes of giant impacts are governed primarily by gravitational forces and angular momentum \citep[e.g.][]{2004IcarusCanup,2010ApJLeinhardt}.

Late-stage giant impacts have already been invoked to explain features of our solar system as far-reaching as the crustal dichotomy of Mars \citep{1984NatureWilhelmsSquyres,2008NatureMarinova}, the high obliquity of Uranus \citep{1992IcarusSlattery,2020MNRASReinhardt,2020IcarusRibeiro}, and the origins of Mercury and the Moon (for reviews, see, e.g., \citealp{2018ApJChau,2014AREPSAsphaug}). While the last giant impact around the Sun was probably 4.3--4.5 billion years ago (perhaps the Moon-forming collision itself), giant impacts in several nearby exoplanetary systems are thought to be recent or ongoing, responsible for observations of late-forming debris rings \citep{2019ApJThompson} or planets with high eccentricity \citep[\num{>0.5};][]{2019ApJLFrelikh}.

\subsection{The diversity of giant impact outcomes}

\begin{figure}
	\centering
	\includegraphics{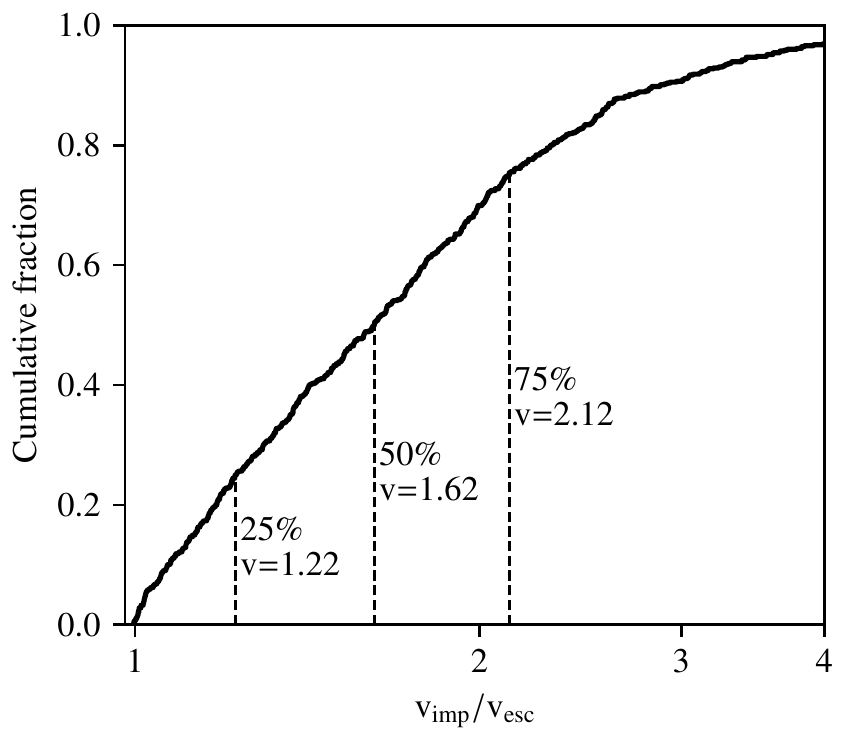}
	\caption{Cumulative distribution of collision velocities (scaled by the mutual escape velocity, given on a logarithmic scale) for collisions on bodies with $m_\mathrm{tar}>\SI{0.1}{\mearth}$ from the \textit{N}-body simulations of \e20 that include the realistic collision model. The locations of the boudaries of four quantiles (with $v_\mathrm{imp}/v_\mathrm{esc}$ given as $v$) are also provided.}
	\label{fig:vel-comp}
\end{figure}

The nature of the giant impact stage, and hence the geology and habitability of planets that are created by this process, depends on several factors. The starting conditions are represented by the orbital distribution of early-forming embryos (a.k.a. oligarchs), and their sizes and compositions are determined by disk physics \citep{2000IcarusKokuboA} and nebula chemistry and transport \citep{2009IcarusCiesla}. Equally important is the velocity distribution of the embryos, which evolves owing to self-stirring or forcing by giant planet perturbations, or damping by planetesimals, or self-regulation by the debris produced by the giant impacts themselves.

Giant impacts occur at velocities that are generally faster than $v_\mathrm{esc}$, because in order for two planets to collide, their orbits must intersect. This requires a relative velocity $v_\mathrm{rel}>0$, where $v_\mathrm{coll}^2=v_\mathrm{esc}^2+v_\mathrm{rel}^2$. In classical self-stirring of a similar-sized population, relative velocities are excited to $\sim v_{\rm esc}$ so that impact velocities of order $1.4 v_{\rm esc}$ are expected \citep{1969BookSafronov}. Considering Earth and Venus formation, for an embryo in circular orbit starting at \SI{1}{\au} to collide with an embryo at \SI{0.7}{\au} requires $v_\mathrm{rel}$ of at least \SI{2.5}{\kilo\meter\per\second}. Lower-velocity collisions near $v_\mathrm{esc}$, including the so-called ``graze-and-merge'' collisions \citep{2010ApJLeinhardt} that lead to successful scenarios for Moon formation \citep[e.g.][]{2001NatureCanup}, are only one outcome among others with similar or greater likelihood \citep{2012ApJLeinhardt}.

Faster giant impacts can also happen, but they are less probable because relative motions are damped by planetesimals and collisions. Also, ejection from the planet formation zone (loss of a participant) becomes more likely as relative velocities increase. Thus, most giant impacts occur in a rather sensitive region of velocity parameter space where accretion may or may not happen. Most \textit{N}-body simulations of terrestrial planet formation indeed find that giant impacts occur in the velocity range $v_\mathrm{coll}/v_\mathrm{esc}\sim 1-2$ with outliers that are faster \citep[e.g.,][]{1999IcarusAgnor,2010ApJKokubo,2016ApJQuintana}.

Figure \ref{fig:vel-comp} shows the distribution of impact velocities for \textit{N}-body simulations in \citet[hereafter \e20]{2020ApJEmsenhuberA}. It should be noted that this set only contains 16 \textit{N}-body simulations and that these all produce planetary systems less massive than the solar system's terrestrial planets. Also, the distribution only accounts for cases where the target mass $m_\mathrm{tar}>\SI{0.1}{\mearth}$ in order to provide a better comparison with the simulations that we present later in this work. The median value of $v_\mathrm{coll}/v_\mathrm{esc}$ is 1.6, which is above the hit-and-run velocity threshold across expected impact angles \citep{2010ApJKokubo,2020ApJGabriel}. These simulations have less dynamical friction than some other studies \citep{2006IcarusRaymond,2006IcarusOBrien,2013IcarusChambers,2019ApJKobayashi} because debris have been ignored, but they have more dynamical friction than studies that assume perfect merging without planetesimals. They have greater gravitational excitation than perfect merging simulations because they allow impact-periapsis `runners' to escape, with the resulting large deflection. Moreover, in our calculations there is no inflation factor, so objects must come within the sum of the radii to detect a collision. The latter aspects tend to increase the rate of collisions early on, as well as their velocities, compared to other studies, but make it harder to ultimately accrete.

The \textit{N}-body simulations by \e20 determine the outcome of an individual giant impact from the key nondimensional parameters, which are the impact angle $\theta_\mathrm{coll}$, scaled velocity $v_\mathrm{coll}/v_\mathrm{esc}$, mass ratio $\gamma=m_\mathrm{imp}/m_\mathrm{tar}$, and composition \citep{2010ChEGAsphaug,2012ApJGenda,2012ApJStewart,2020ApJGabriel} represented by the metallic core radius relative to the planet radius. For a differentiated chondritic sphere with \SI{\sim30}{wt\percent} iron and \SI{70}{wt\percent} silicate mantle the core is half the planet's radius.

Initial rotation, normalized to the spin-disruption limit, is another important parameter, and while it has been included in prior studies \citep{2008IcarusCanup} and special cases close to the breakup limit \citep{2012ScienceCuk}, the free parameters add up quickly, especially in three dimensions, making it difficult to run a systematic study at adequate numerical resolution. The N-body simulations of \e20 are therefore based on the assumption that the bodies are not spinning prior to each collision, and rotation is ignored.

To obtain a sufficiently accurate working model of giant impacts, for the regime of nonrotating chondritic bodies and in the mass range of \num{e-3} and \SI{1}{\mearth}, the \textit{N}-body simulations in \e20 used the surrogate model of giant impacts \citep[][hereafter \c19]{2019ApJCambioni}. This is a set of machine-learning algorithms trained on 800 smoothed particle hydrodynamics (SPH) simulations of giant impacts, varying the impact angle, velocity, and mass ratio within the limits defined by the ranges explored in building the dataset. Sparsely sampled parameter spaces such as these can be used by machine learning to generalize the underlying model. \c19 have used machine learning to classify the outcomes (disruption, accretion, graze-and-merge, hit-and-run) and to develop a neural network that accurately predicts the mass of the largest remnant $m_\mathrm{lr}$, or specifically the accretion efficiency
\begin{equation}
\xi=(m_\mathrm{lr}-m_\mathrm{tar})/m_\mathrm{imp}
\end{equation}
within the parameter limits of the database. \e20 have also developed neural networks in the case of hit-and-runs that predict the masses and velocity vectors of the target and runner.
As mentioned above, \e20 have not included the fate and influence of debris produced by giant impacts, which are defined as anything smaller than the runner, objects that are generally orders of magnitude lower in mass. Over the course of the late stage, the cumulative debris can represent a substantial fraction of the total mass (\e20) and reduce the eccentricities and inclinations of the embryos.

\subsection{The perfect merging assumption}

From a perspective of planetary formation, the variety of giant impact outcomes described in the previous section would not matter so much if the leftover material is reaccreted later by the same target. This assumption can seem justifiable, because whenever two planets emerge from a hit-and-run collision they may be expected to experience a follow-on collision, so that merger seems to be a foregone conclusion. Most \textit{N}-body simulations of terrestrial planet formation \citep[e.g.][]{1999MNRASChambers,2006IcarusOBrien,2014EPSLFischerCiesla} have traditionally used this approach and treated impacts as perfect mergers, $\xi=1$ for collisions that are not catastrophically disruptive.

However, \citet[hereafter \paperone][]{2019ApJEmsenhuberA} showed that this is not generally the case. They studied the fate of the runner following hit-and-runs into proto-Earths at \SI{1}{\au}, for thousands of geometries, and found that, contrary to expectation, only about half the time (depending on the runner's egress velocity, which depends on the impact velocity and angle) do they return to collide again with proto-Earth. When they do, the return collision happens on a timescale of thousands to millions of years.

In this sense, the assumption of perfect merging would lead to a poor estimate of the accretion timescale, and the inferred thermodynamic (e.g., lack of cooling between collisions in a chain) and differentiation history \citep{2021PSJCambioni} would be unrealistic. Furthermore, \paperone{} found that a majority of those runners that do not return to proto-Earth are likely to collide with Venus, assuming the present masses and orbits of the planets. They also showed that for returning runners, the impact velocity of the second collision is usually similar to the egress velocity following the hit-and-run, which is slower than the original impact owing to momentum loss. So, the follow-on collisions tend to be comparatively slower than the previous hit-and-run encounter. The offset angle between impacts, however, is uncorrelated and random, meaning that the returning collision is off-axis by about \ang{90} on average.

\subsection{Fate of the runner}
\label{sec:hitrun}

In most hit-and-runs, a large part of the projectile survives the giant impact. The velocity of the runner can be strongly reduced in magnitude and deflected in direction relative to the inbound velocity \citep{2012ApJGenda}. If the runner is slowed so that the two bodies are bound gravitationally after the collision, then instead of a hit-and-run it is a graze-and-merge collision \citep[e.g.,][]{2010ApJLeinhardt}, the archetype being the impact origin of the Moon \citep{2001NatureCanup}. The boundary between hit-and-run and graze-and-merge occurs when the runner gets to about two-thirds of the Hill radius, as studied by \citet{2019ApJEmsenhuberB}. This leads to a sensitivity around the boundary between hit-and-run and merger.

Hit-and-runs can be further subdivided into low-velocity hit-and-runs (that is, not much faster than the escape velocity), where the runner is an identifiable remnant of the projectile (e.g., a mantle-stripped core, sometimes barely so) and high-velocity hit-and-runs where the runner is disrupted or dispersed \citep{2006NatureAsphaug}. Even in dispersive hit-and-runs the target remains mostly intact; disruption of the target by giant impact requires much greater energies than are considered here, 3 to 5~$v_\mathrm{esc}$ or more. We ignore these faster collisions for now, for several reasons: they are much less probable \citep[Figure~\ref{fig:vel-comp} of this work and, e.g., ][]{1999IcarusAgnor}, their debris reaccumulation is not a single event and thus harder to constrain, and the surrogate model can be used with highest fidelity at velocities lower than 3~$v_\mathrm{esc}$.

\paperone{} modeled a series of dynamical evolutions of the two largest remnants (target and runner) after relatively low velocity hit-and-runs with the proto-Earth, an \SI{0.9}{\mearth} planet at \SI{1}{\au}. The authors found the following:
\begin{itemize}
    \item Between one-third and two-thirds of the runners collide back onto proto-Earth, depending on the egress velocity, after an interlude of thousands to millions of years.
    \item Of those that do collide with another body, most end up on Venus. Indeed, it was found that the destination of moderate-velocity hit-and-runs with proto-Earth is almost as likely to be Venus as proto-Earth itself.
    \item Venus-bound runners following hit-and-runs with proto-Earth have impact velocities that are faster than their counterparts having return collisions with proto-Earth, implying that accretion might not end with this event, leading to longer collision chains.
\end{itemize}

This motivates us to explore more generally the fate of runners and their exchanges between Venus and the Earth, where our aim is to identify systematic differences in the kinds of late-stage collisions they each experience, which might account for the major differences in the geophysical and dynamical states of these otherwise-similar ``sister planets.''

\subsection{This work}
\label{sec:thiswork}

Here we explore how the capture, loss, and interchange of runners affects the growth of Venus compared with the Earth. Collision chains start with a hit-and-run, potentially involve multiple subsequent hit-and-runs, and end in terminal mergers (\paperone). To study the general process, we need to model each collision, especially in the case where there may be one or more intermediate hit-and-runs. To make this analysis tractable, we cannot simulate every giant impact explicitly, and thus we utilize the work of \e20 who extended the machine-learning approach, training on the same simulation data as \c19, \citet{2011PhDReufer}, and \citet{2020ApJGabriel} to retrieve not only the accretion efficiency (mass added to or removed from the target) but also the properties of the two main remnants, in the case of hit-and-run, and their relative orbits (i.e. egress velocities). The result is a machine-learning-derived surrogate model for giant impacts, providing a functional map of input parameters into outputs that allows us to realistically model collisions as they happen during the \textit{N}-body evolution.

Using this machine-learning surrogate model, it is possible to perform \textit{N}-body evolution that can continue from the initial hit-and-run through the subsequent collisions, while improving the fidelity in the dynamics of the system compared to other models. This allows us to extend the procedure of \paperone{} to continue the dynamical evolution past the first subsequent collision, to model collision chains lasting one, two, three, or more hit-and-runs until the terminal merger, using the procedure described in \c19 and \e20. We are able to determine the ultimate destination of the runner and the number of intermediate collisions required until final accretion.

\section{Methods}

Our procedure is similar to \paperone{}, but several improvements have been implemented. The first collision of the chain (C\textsubscript{1}) is modeled directly with SPH to simulate several candidate low-velocity hit-and-runs into proto-Earth and proto-Venus at high resolution. The results (masses and velocities of the two largest remnants) are then transferred into an \textit{N}-body code to track the dynamical evolution of proto-Earth or proto-Venus and the runner.

For each first collision, we perform \num{1000} \textit{N}-body evolutions for random collision orientations to obtain a statistical description of the fate of the runner. During the \textit{N}-body evolution, subsequent collisions (C\textsubscript{2}\ldots C\textsubscript{F}, for collisions 2 through F) are treated using surrogate models following the methodology of \e20. Each \textit{N}-body evolution ends when the runner is lost (by accretion onto another body or ejection) or the maximum time elapses (generally \SI{50}{\mega\year}, but extended to \SI{400}{\mega\year} for some cases in Section~\ref{sec:res-time}).

\subsection{Initial collision simulations}

To model the initial collisions, we use an SPH code that has been developed for modeling similar-sized collisions in the size and velocity regime of late-stage terrestrial planet formation (\citealp{2012IcarusReufer,2018IcarusEmsenhuber}; \paperone), with targets of \SI{\sim0.1}{\mearth} and larger. For these size bodies, strength effects can be ignored, so the bodies are treated as fluids. SPH is a Lagrangian technique with material subdivided into mass particles. A kernel interpolation is used to compute hydrodynamic quantities at any location. Spatial derivatives are computed using an interpolation with the derivatives of the kernel. Time evolution is provided by the Euler equations, except for the density, which is retrieved using the kernel interpolation with a correction term for particles close to the surface \citep{2017MNRASReinhardtStadel}. The pressure $p(\rho,S)$ and the other physical quantities necessary for the hydrodynamical equations are retrieved using a tabulated form of the M-ANEOS equation of state \citep{ANEOS,2007M&PSMelosh}. To compute self-gravity and retrieve neighboring particles, a hierarchical spatial tree is used \citep{1986NatureBarnesHut}. For reviews on SPH, the reader is referred to, e.g., \citet{1992ARA&AMonaghan} and \citet{2009NARRosswog}.

\subsubsection{Thermodynamic equation}

Energy conservation in the Euler equations is provided by the energy equation. In its standard form, the equation is simply the adiabatic compression or expansion of the material. In SPH, it is modified (as is the momentum equation) for the addition of the artificial viscosity to handle shocks (see below). Nevertheless, in fluid SPH (i.e., without additional solid forces) the largest contribution to the changes in internal energy comes from adiabatic compression or expansion.

To avoid the numerical integration of the energy equation for an adiabatic process, which can lead to nonphysical states \citep{2017MNRASReinhardtStadel}, we use the first law of thermodynamics to determine an entropy equation,
\begin{equation}
\frac{\partial S}{\partial t}=\frac{1}{T}\left(\frac{\partial u}{\partial t}+p\frac{\partial V}{\partial t}\right)=\frac{1}{T}\frac{\partial u_\mathrm{AV}}{\partial t},
\end{equation}
where $S$ is the specific entropy, $T$ is the temperature, $u$ is the specific internal energy, $p$ is the pressure, $V=1/\rho$ is the specific volume, and the $\partial u_\mathrm{AV}/\partial t$ term denotes the contribution of artificial viscosity to the energy equation. A drawback of this formulation is that the integration of the entropy equation introduces additional imprecision for the artificial viscosity. This can lead to the total energy suffering from a slightly larger drift overall, which is not expected to affect the results substantially.

\subsubsection{Artificial viscosity}

Artificial viscosity is a nearly ubiquitous numerical tool for resolving shocks in SPH. The standard formulation that was used in previous studies suffers from drawbacks, such as being triggered in cases where no shocks occur, for instance, in shear motions.
Here, we use a form that is inspired by Riemann solvers \citep{1997JCPMonaghan}. The acceleration term is computed as
\begin{equation}
    \frac{\partial \mathbf{v}_\mathrm{AV,i}}{\partial t}=\sum_\mathrm{j}\frac{m_\mathrm{j}}{\rho_\mathrm{ij}}\alpha_\mathrm{ij}v_\mathrm{sig}\left(\mathbf{v}_\mathrm{ij}\cdot\hat{r}_\mathrm{ij}\right)\nabla_\mathrm{i}W_\mathrm{ij},
\end{equation}
where $i$ and $j$ are particle indexes, $m_\mathrm{j}$ is the particle mass, $\rho_\mathrm{ij}=(\rho_\mathrm{i}+\rho_\mathrm{j})/2$ is the averaged particle density, $\mathbf{v}_\mathrm{ij}=\mathbf{v}_\mathrm{i}-\mathbf{v}_\mathrm{j}$ the relative velocity of the particles, $\hat{r}_\mathrm{ij}=\mathbf{r}_\mathrm{ij}/r_\mathrm{ij}$ with $r_\mathrm{ij}=\mathbf{r}_\mathrm{i}-\mathbf{r}_\mathrm{j}$ the unit vector along the relative position of the two particles, and $\nabla_\mathrm{i}W_\mathrm{ij}$ is the gradient of the kernel $W(r_\mathrm{ij},h_\mathrm{ij})$ with respect to $r_\mathrm{i}$. $v_\mathrm{sig}$ represents the maximum signal velocity between particles $i$ and $j$ and can be estimated as
\begin{equation}
    v_\mathrm{sig}=c_\mathrm{s,i}+c_\mathrm{s,j}-\mathbf{v}_\mathrm{i,j}\cdot\hat{r}_\mathrm{i,j},
\end{equation}
where $c_\mathrm{s}$ is the sound speed from the equation of state \citep{1997JCPMonaghan}.

The corresponding energy change is given by
\begin{equation}
    \frac{\partial u_\mathrm{AV,i}}{\partial t}=\sum_\mathrm{j}\frac{m_\mathrm{j}}{2\rho_\mathrm{ij}} \alpha_\mathrm{ij}v_\mathrm{sig}\left(\mathbf{v}_\mathrm{ij}\cdot\hat{r}_\mathrm{ij}\right)^2\left(\nabla_\mathrm{i}W_\mathrm{ij}\cdot\hat{r}_\mathrm{ij}\right).
\end{equation}

In SPH simulations, artificial viscosity can trigger in unwarranted circumstances, e.g., during shear motion in accretion disks and potentially in hit-and-runs. A simple method to reduce the artificial viscosity is the ``Balsara switch'' \citep{1991PhDBalsara} to apply artificial viscosity only during compression. This can, however, cause problems with shocks in accretion disks \citep{2004JCPOwen} and artificial particle alignments. So, in our case, we have decided instead to use time-dependent viscosity factors following \citet{1997JCPMorrisMonaghan}. Here the $\alpha$ parameter of the artificial viscosity is treated as another time-integrated quantity of each particle. Its derivative is given by
\begin{equation}
    \frac{\partial\alpha_\mathrm{i}}{\partial t}=\max{\left(-(\nabla\cdot\mathbf{v})_\mathrm{i}\left(\alpha_\mathrm{max}-\alpha_\mathrm{i}\right),0\right)}-\frac{\alpha_\mathrm{i}-\alpha_\mathrm{min}}{\tau_\mathrm{i}},
\end{equation}
with
\begin{equation}
    \tau_\mathrm{i}=\frac{h_\mathrm{i}}{\xi c_\mathrm{s,i}}
\end{equation}
being the timescale over which $\alpha_\mathrm{i}$ is reduced and $\xi=0.1$. This is set so that the $\alpha_\mathrm{i}$ factor is reduced after shock passed by about 10 smoothing length. In our case, we set $\alpha_\mathrm{min}=0.1$ and $\alpha_\mathrm{max}=1.5$.

\subsubsection{Initialization}

We construct the projectile and target bodies using an improved methodology over that of \paperone{}.
The first step is to obtain a 1D radial hydrostatic profile using the scheme of \citet{1991LNPBenz}. With this profile an initial SPH body is generated using the methodology described in \citet{2017MNRASReinhardtStadel}. This body is made of successive layers of SPH particles, whose properties are taken from the 1D radial profile. The locations of the particles on each layer are obtained using the \texttt{HEALPix} \citep{2005ApJGorski} software package. The location of SPH particles in each layer is found iteratively with the goal to minimize the difference of horizontal and vertical spacing between the particles. The resolution is chosen so that the number of particles is proportional to the body's mass, with a \SI{1}{\mearth} body being represented by \num{500000} SPH particles. This results in initial SPH bodies whose particles have a lower residual velocity. Nevertheless, we further evolve these bodies for \SI{6}{\hour} of physical time, which further reduces the rms of the residual velocities to about \SI{2.5}{\meter\per\second}, or \num{2.5e-4} of the escape velocity in our Earth-scale planets.

\subsection{Evolution until Subsequent Collision}

The largest remnants of the hit-and-runs are identified following the SPH simulation using the methodology of \e20. These bodies, the target and runner, are
then mapped into the \texttt{mercury6} \textit{N}-body code \citep{1999MNRASChambers} to follow their dynamical evolution. We include the evolution of the other major planets as well, assuming their present orbits, which is important considering our result that planets can often exchange runners. The presence of Jupiter and Saturn on their present-day orbits is consistent with them either forming eccentric \citep[e.g.,][]{2021ApJWoo} or having migrated early in the process of terrestrial planet formation \citep[e.g.,][]{2018IcarusClement}.

For now, the lesser debris produced by the hit-and-run is ignored in the further evolution, and in any case for our simulations at most \SI{2.4}{\percent} of the material is in objects smaller than the runner ($m_\mathrm{lost}$ as reported in Tables~\ref{tab:res-coll-earth} and \ref{tab:res-coll-venus}). We proceed as in \paperone: for each hydrodynamical simulation, we perform \num{1000} realizations of the dynamical evolution, each with a different orientation of the pre-impact orbit of the impactor. The target is assumed to be on a circular orbit, while the orbit of the impactor is computed according to \citet{2018MNRASJackson}. To obtain the orientation, we use a Monte Carlo approach, and we assume that the orientation of the relative orbit follows a uniform distribution in space (\paperone).

The criterion for the end of the dynamical evolution has been updated following the improvement in our treatment of collision handling in the \textit{N}-body evolution, described below. Using surrogate models for giant impacts, we are now able to continue the dynamical evolution until the runner has been fully accreted, which can require successive collisions, or until a predetermined time that we set to \SI{50}{\mega\year}, as justified in Section~\ref{sec:res-time}, with several cases extended further to \SI{400}{\mega\year}.

\subsubsection{Collision model}

The underlying methodology of the collisional handling during \textit{N}-body evolution is described in \c19. A data set comprising about 800 SPH simulations that were obtained by \cite{2011PhDReufer} was used to performed machine learning (data available in \citealp{2020ApJGabriel}). The simulations span a range of target masses $m_\mathrm{tar}$ from \num{e-2} to \SI{1}{\mearth}, impactor mass ratio $\gamma=m_\mathrm{imp}/m_\mathrm{tar}$ from 0.1 to 0.7, and impact velocity ratio $v_\mathrm{coll}/v_\mathrm{esc}$ from 1 to 4. This dataset is well suited for this study, as it encompasses the parameter range of targets and runners from our SPH simulations. A classifier was trained on the dataset to determine whether a collision is in the hit-and-run regime (two remnants) or not (single remnant). Then, two neural networks were also trained to obtain one regressor that provides the mass of the two largest remnants and another regressor that provides the orbital characteristics of the two remnants in the case of a hit-and-run (\e20). The classifier and neural networks were implemented in the \texttt{collresolve}\footnote{\url{https://github.com/aemsenhuber/collresolve}} library \citep{2019SoftwareEmsenhuberCambioni} that we are using in this work. The full procedure for the treatment of collisions and its adaptation to the \texttt{mercury6} code are discussed in detail in \e20.

Additionally, we determine the mass exchange between the two bodies in a similar way that is computed for the Earth-disk equilibration in simulation of potential Moon-forming giant impacts \citep{2012IcarusReufer}:
\begin{equation}
    \delta f_\mathrm{T} = \frac{f^\mathrm{s\leftarrow t}_\mathrm{mant}}{f^\mathrm{l\leftarrow t}_\mathrm{mant}}-1 = \frac{f^\mathrm{s\leftarrow t}_\mathrm{mant}}{1-f^\mathrm{l\leftarrow i}_\mathrm{mant}}-1,
    \label{eq:dft}
\end{equation}
where $f^\mathrm{l\leftarrow t}_\mathrm{mant}$ and $f^\mathrm{s\leftarrow t}_\mathrm{mant}$ are the mass fraction of the mantle of largest and second remnants that are coming from the target, respectively, and $f^\mathrm{l\leftarrow i}_\mathrm{mant}$ is the mantle mass fraction of the largest remnant coming from the impactor. These results for mass exchange are not important for the present study but are applied to the third paper in this series, on the origin of the Moon \citep{2021PSJAsphaug}.

A value of \SI{-100}{\percent} indicates that the mantles do not mix at all during the collision, i.e., the mantle of the largest remnant is made exclusively of target material, while the mantle of the second remnant is made exclusively of impactor material. A value of \SI{0}{\percent} indicates that each remnant has a mantle made of the same proportion of target and impactor material. We assume that this exchanged material becomes well mixed after the collision, within each core and mantle, respectively. The collisions where the two remnants remain gravitationally bound after the collision (graze-and-merge) are analyzed using the procedure outlined in \citet{2019ApJEmsenhuberB}.

\section{Results}
\label{sec:results}

\subsection{Initial Hit-and-run Collisions with Proto-Earth}
\label{sec:res-initial}

For the SPH simulations of the initial hit-and-runs with the proto-Earth, we start with bodies whose masses are $m_\mathrm{tar}=\SI{0.9}{\mearth}$ and $m_\mathrm{imp}=\SI{0.15}{\mearth}$. The impactor is somewhat smaller than in \paperone; this is to end up with an approximately Mars-sized runner, which would be the suitable size for the scenarios of Moon formation \citep{2017GRLPiet}, which we consider in the next paper in this series \citep{2021PSJAsphaug}.
In addition, we perform new SPH simulations of hit-and-runs more relevant to Venus, with $m_\mathrm{tar}=\SI{0.7}{\mearth}$ and $m_\mathrm{imp}=\SI{0.15}{\mearth}$.

\begin{table*}
    \begin{center}
    \caption{Outcome of SPH simulations for collisions with Earth.}
    \begin{tabular}{ccccccccccccccc}
        \hline
        $\frac{v_\mathrm{coll}}{v_\mathrm{esc}}$ & $\theta_\mathrm{coll}$ & $m_\mathrm{lr}$ & $m_\mathrm{sr}$ & $m_\mathrm{lost}$ & $\frac{v_\mathrm{dep}}{v_\mathrm{esc}}$ & $\theta_\mathrm{dep}$ & $P_\mathrm{lr}$ & $Z_\mathrm{lr}$ & $Z_\mathrm{sr}$ & $f^\mathrm{l\leftarrow i}_\mathrm{core}$ & $f^\mathrm{s\leftarrow t}_\mathrm{core}$ & $f^\mathrm{l\leftarrow i}_\mathrm{mant}$ & $f^\mathrm{s\leftarrow t}_\mathrm{mant}$ & $\delta f_\mathrm{T}$ \\
        & [deg] & [\si{\mearth}] & [\si{\mearth}] & [\si{\mearth}] & & [deg] & [hr] \\
        \hline
        \num{1.10} & \num{52.5} & \num{0.93} & \num{0.12} & \num{2.5e-4} & \num{1.00} & \num{58.9} & \num{13.2} & \SI{29.5}{\percent} & \SI{33.8}{\percent} & \SI{1.7}{\percent} & \SI{0.0}{\percent} & \SI{5.2}{\percent} & \SI{10.4}{\percent} & \SI{-89.1}{\percent} \\
        \num{1.10} & \num{55.0} & \num{0.92} & \num{0.13} & \num{1.3e-4} & \num{1.01} & \num{59.6} & \num{15.1} & \SI{29.5}{\percent} & \SI{33.3}{\percent} & \SI{0.9}{\percent} & \SI{0.0}{\percent} & \SI{4.2}{\percent} & \SI{9.1}{\percent} & \SI{-90.5}{\percent} \\
        \num{1.10} & \num{60.0} & \num{0.91} & \num{0.14} & \num{5.1e-5} & \num{1.04} & \num{62.3} & \num{19.5} & \SI{29.5}{\percent} & \SI{33.1}{\percent} & \SI{0.1}{\percent} & \SI{0.0}{\percent} & \SI{3.0}{\percent} & \SI{5.7}{\percent} & \SI{-94.1}{\percent} \\
        \num{1.15} & \num{47.5} & \num{0.94} & \num{0.11} & \num{9.6e-4} & \num{1.01} & \num{56.5} & \num{11.0} & \SI{29.8}{\percent} & \SI{32.2}{\percent} & \SI{3.3}{\percent} & \SI{0.0}{\percent} & \SI{6.2}{\percent} & \SI{15.4}{\percent} & \SI{-83.6}{\percent} \\
        \num{1.15} & \num{50.0} & \num{0.93} & \num{0.12} & \num{5.6e-4} & \num{1.03} & \num{57.7} & \num{12.5} & \SI{29.7}{\percent} & \SI{32.6}{\percent} & \SI{2.1}{\percent} & \SI{0.0}{\percent} & \SI{5.2}{\percent} & \SI{12.8}{\percent} & \SI{-86.5}{\percent} \\
        \num{1.15} & \num{55.0} & \num{0.92} & \num{0.13} & \num{3.5e-4} & \num{1.07} & \num{60.4} & \num{16.2} & \SI{29.5}{\percent} & \SI{33.3}{\percent} & \SI{0.6}{\percent} & \SI{0.0}{\percent} & \SI{3.8}{\percent} & \SI{7.9}{\percent} & \SI{-91.7}{\percent} \\
        \num{1.15} & \num{60.0} & \num{0.91} & \num{0.14} & \num{2.3e-4} & \num{1.09} & \num{62.5} & \num{21.1} & \SI{29.6}{\percent} & \SI{32.8}{\percent} & \SI{0.0}{\percent} & \SI{0.0}{\percent} & \SI{2.7}{\percent} & \SI{5.0}{\percent} & \SI{-94.9}{\percent} \\
        \num{1.20} & \num{42.5} & \num{0.95} & \num{0.08} & \num{2.5e-2} & \num{0.99} & \num{37.9} & \num{9.3} & \SI{30.2}{\percent} & \SI{37.1}{\percent} & \SI{5.7}{\percent} & \SI{0.0}{\percent} & \SI{7.2}{\percent} & \SI{23.3}{\percent} & \SI{-74.9}{\percent} \\
        \num{1.20} & \num{43.0} & \num{0.95} & \num{0.08} & \num{2.0e-2} & \num{1.00} & \num{42.1} & \num{9.5} & \SI{30.1}{\percent} & \SI{36.6}{\percent} & \SI{5.4}{\percent} & \SI{0.0}{\percent} & \SI{7.3}{\percent} & \SI{22.8}{\percent} & \SI{-75.4}{\percent} \\
        \num{1.20} & \num{45.0} & \num{0.94} & \num{0.11} & \num{2.4e-3} & \num{1.04} & \num{52.1} & \num{10.4} & \SI{30.0}{\percent} & \SI{31.1}{\percent} & \SI{4.0}{\percent} & \SI{0.0}{\percent} & \SI{6.4}{\percent} & \SI{17.9}{\percent} & \SI{-80.8}{\percent} \\
        \num{1.20} & \num{50.0} & \num{0.93} & \num{0.12} & \num{1.7e-3} & \num{1.09} & \num{56.1} & \num{13.4} & \SI{29.7}{\percent} & \SI{32.9}{\percent} & \SI{1.6}{\percent} & \SI{0.0}{\percent} & \SI{4.7}{\percent} & \SI{11.3}{\percent} & \SI{-88.1}{\percent} \\
        \num{1.20} & \num{55.0} & \num{0.92} & \num{0.13} & \num{1.0e-3} & \num{1.12} & \num{58.5} & \num{17.4} & \SI{29.5}{\percent} & \SI{33.3}{\percent} & \SI{0.3}{\percent} & \SI{0.0}{\percent} & \SI{3.4}{\percent} & \SI{6.9}{\percent} & \SI{-92.9}{\percent} \\
        \hline
    \end{tabular}
    \tablecomments{$v_\mathrm{coll}/v_\mathrm{esc}$ and $\theta_\mathrm{coll}$ are the initial conditions, while the other columns are the results. $m_\mathrm{lr}$, $m_\mathrm{sr}$, and $m_\mathrm{lost}$ are the mass of the largest remnant, second remnant, and debris. $v_\mathrm{dep}/v_\mathrm{esc}$ and $\theta_\mathrm{dep}$ are computed by analogy to the initial conditions, but refer to the orbits of the bodies after the collision. $P_\mathrm{lr}$ is the rotation period of the largest remnant, $Z_\mathrm{lr}$ is its core mass fraction, while $Z_\mathrm{sr}$ is the core mass fraction of the second remnant. $f^\mathrm{l\leftarrow i}_\mathrm{core}$ and $f^\mathrm{s\leftarrow t}_\mathrm{core}$ are the core mass fraction of the largest and second remnants coming from the impactor and target, respectively. $f^\mathrm{l\leftarrow i}_\mathrm{mant}$ and $f^\mathrm{s\leftarrow t}_\mathrm{mant}$ are similar, but for the mantle. $\delta f_\mathrm{T}$ is the mantle equilibration factor, Eq.~(\ref{eq:dft}).}
    \label{tab:res-coll-earth}
    \end{center}
\end{table*}

\begin{table*}
    \begin{center}
    \caption{Outcome of SPH simulations for collisions with Venus.}
    \begin{tabular}{ccccccccccccccc}
        \hline
        $\frac{v_\mathrm{coll}}{v_\mathrm{esc}}$ & $\theta_\mathrm{coll}$ & $m_\mathrm{lr}$ & $m_\mathrm{sr}$ & $m_\mathrm{lost}$ & $\frac{v_\mathrm{dep}}{v_\mathrm{esc}}$ & $\theta_\mathrm{dep}$ & $P_\mathrm{lr}$ & $Z_\mathrm{lr}$ & $Z_\mathrm{sr}$ & $f^\mathrm{l\leftarrow i}_\mathrm{core}$ & $f^\mathrm{s\leftarrow t}_\mathrm{core}$ & $f^\mathrm{l\leftarrow i}_\mathrm{mant}$ & $f^\mathrm{s\leftarrow t}_\mathrm{mant}$ & $\delta f_\mathrm{T}$ \\
        & [deg] & [\si{\mearth}] & [\si{\mearth}] & [\si{\mearth}] & & [deg] & [hr] \\
        \hline
        \num{1.10} & \num{52.5} & \num{0.72} & \num{0.13} & \num{1.8e-4} & \num{0.99} & \num{53.4} & \num{12.7} & \SI{29.5}{\percent} & \SI{32.9}{\percent} & \SI{1.2}{\percent} & \SI{0.0}{\percent} & \SI{5.5}{\percent} & \SI{11.4}{\percent} & \SI{-87.9}{\percent} \\
        \num{1.10} & \num{55.0} & \num{0.72} & \num{0.13} & \num{4.8e-5} & \num{1.01} & \num{59.4} & \num{14.0} & \SI{29.4}{\percent} & \SI{33.3}{\percent} & \SI{0.5}{\percent} & \SI{0.0}{\percent} & \SI{4.9}{\percent} & \SI{8.7}{\percent} & \SI{-90.9}{\percent} \\
        \num{1.10} & \num{60.0} & \num{0.71} & \num{0.14} & \num{1.6e-5} & \num{1.03} & \num{61.4} & \num{17.4} & \SI{29.5}{\percent} & \SI{32.6}{\percent} & \SI{0.0}{\percent} & \SI{0.0}{\percent} & \SI{3.5}{\percent} & \SI{6.0}{\percent} & \SI{-93.8}{\percent} \\
        \num{1.15} & \num{47.5} & \num{0.73} & \num{0.12} & \num{4.3e-4} & \num{1.00} & \num{54.9} & \num{10.8} & \SI{29.6}{\percent} & \SI{32.8}{\percent} & \SI{2.6}{\percent} & \SI{0.0}{\percent} & \SI{6.9}{\percent} & \SI{14.1}{\percent} & \SI{-84.8}{\percent} \\
        \num{1.15} & \num{50.0} & \num{0.72} & \num{0.13} & \num{2.5e-4} & \num{1.03} & \num{55.2} & \num{12.0} & \SI{29.5}{\percent} & \SI{32.7}{\percent} & \SI{1.6}{\percent} & \SI{0.0}{\percent} & \SI{5.8}{\percent} & \SI{12.4}{\percent} & \SI{-86.8}{\percent} \\
        \num{1.15} & \num{55.0} & \num{0.72} & \num{0.13} & \num{1.4e-4} & \num{1.06} & \num{60.5} & \num{15.0} & \SI{29.5}{\percent} & \SI{32.9}{\percent} & \SI{0.3}{\percent} & \SI{0.0}{\percent} & \SI{4.3}{\percent} & \SI{8.2}{\percent} & \SI{-91.5}{\percent} \\
        \num{1.15} & \num{60.0} & \num{0.71} & \num{0.14} & \num{9.9e-5} & \num{1.08} & \num{60.5} & \num{19.0} & \SI{29.6}{\percent} & \SI{32.6}{\percent} & \SI{0.0}{\percent} & \SI{0.0}{\percent} & \SI{3.1}{\percent} & \SI{5.5}{\percent} & \SI{-94.3}{\percent} \\
        \num{1.20} & \num{42.5} & \num{0.74} & \num{0.11} & \num{1.9e-3} & \num{1.00} & \num{51.2} & \num{9.4} & \SI{29.8}{\percent} & \SI{32.0}{\percent} & \SI{4.9}{\percent} & \SI{0.0}{\percent} & \SI{8.6}{\percent} & \SI{18.8}{\percent} & \SI{-79.5}{\percent} \\
        \num{1.20} & \num{45.0} & \num{0.73} & \num{0.12} & \num{9.3e-4} & \num{1.03} & \num{53.2} & \num{10.3} & \SI{29.8}{\percent} & \SI{31.6}{\percent} & \SI{3.4}{\percent} & \SI{0.0}{\percent} & \SI{7.1}{\percent} & \SI{16.8}{\percent} & \SI{-81.9}{\percent} \\
        \num{1.20} & \num{50.0} & \num{0.72} & \num{0.13} & \num{6.7e-4} & \num{1.08} & \num{56.0} & \num{12.7} & \SI{29.5}{\percent} & \SI{32.8}{\percent} & \SI{1.1}{\percent} & \SI{0.0}{\percent} & \SI{5.4}{\percent} & \SI{11.1}{\percent} & \SI{-88.2}{\percent} \\
        \num{1.20} & \num{55.0} & \num{0.71} & \num{0.14} & \num{4.6e-4} & \num{1.12} & \num{59.7} & \num{15.8} & \SI{29.5}{\percent} & \SI{32.8}{\percent} & \SI{0.1}{\percent} & \SI{0.0}{\percent} & \SI{4.0}{\percent} & \SI{7.2}{\percent} & \SI{-92.5}{\percent} \\
        \hline
    \end{tabular}
    \tablecomments{Columns are the same as in Table~\ref{tab:res-coll-earth}.}
    \label{tab:res-coll-venus}
    \end{center}
\end{table*}

The results are provided in Table~\ref{tab:res-coll-earth} for the collisions with the proto-Earth and in Table~\ref{tab:res-coll-venus} for those with proto-Venus. All the initial bodies are nonrotating, which was chosen for two reasons. First, the most common orientation is for impact orientation and spin to be perpendicular, which provide little angular moment. The second reason is that pre-impact spin affects the outcomes less than the other effects we study here \citep{2020CACTimpe}. We provide several new outputs of the simulations compared to \paperone{} that are related to the SPH modeling of possible return collisions. $P_\mathrm{lr}$ is the rotation period of the largest remnant, computed as $P_\mathrm{lr}=2\pi/\omega_\mathrm{lr}$, with $\omega_\mathrm{lr}=L_\mathrm{lr}/I_\mathrm{lr}$, $L_\mathrm{lr}$ the spin angular momentum, and $I_\mathrm{lr}$ the moment of inertia along the direction of $L_\mathrm{lr}$. Also, $Z_\mathrm{lr}$ and $Z_\mathrm{sr}$ are the core mass fractions of the largest and second remnants.

For the new analyzed quantities from the SPH collisions (i.e., $P_\mathrm{lr}$ and $\delta f_\mathrm{T}$), we see that the rotation period and mixing factor of the largest remnant are mostly dependent on the impact angle, with limited effect of the velocity. It should be noted that all the rotation is induced by the collision as the initial bodies are nonrotating. The core mass fraction of the largest remnant is not much affected by the collision, compared to the original target. It usually slightly decreases because there is often some accretion of impactor mantle material in a hit-and-run \citep{2020ApJGabriel}. In some cases, though, for common impact angles (\ang{42.5} and \ang{43.0} at $v_\mathrm{coll}/v_\mathrm{esc}=1.20$ in the case of collisions with proto-Earth) it increases by a very small amount. The runner always ends up with a higher core mass fraction than the impactor, which is consistent with previous simulations \citep{2014NatGeoAsphaug,2018ApJChau} and planetary differentiation studies \citep{2021PSJCambioni}. Usually, steeper impact angles result in the largest core mass fractions, but this is not always the case. For instance, for collisions with the proto-Earth at $v_\mathrm{coll}/v_\mathrm{esc}=1.20$, the lowest core mass fraction is found for the collision at \ang{45}; the core mass fraction increases again in the simulations with \ang{50} and \ang{55}.

\subsection{The ultimate destination of the runner}

\begin{table*}
    \begin{center}
    \caption{Outcome of dynamical evolution for collisions with Earth.}
    \begin{tabular}{|cc|rrrrrr|rrrrrr|}
        \hline
        $\frac{v_\mathrm{coll}}{v_\mathrm{esc}}$ & $\theta_\mathrm{coll}$ & $N_\mathrm{None}^2$ & $N_\mathrm{Merc}^2$ & $N_\mathrm{Venus}^2$ & $N_\mathrm{Earth}^2$ & $N_\mathrm{Mars}^2$ & $N_\mathrm{Jup}^2$ & $N_\mathrm{None}^\mathrm{F}$ & $N_\mathrm{Merc}^\mathrm{F}$ & $N_\mathrm{Venus}^\mathrm{F}$ & $N_\mathrm{Earth}^\mathrm{F}$ & $N_\mathrm{Mars}^\mathrm{F}$ & $N_\mathrm{Jup}^\mathrm{F}$ \\
        & [deg] & & & & & & \num{} & & & & & & \\
        \hline
        \hline
        \num{1.10} & \num{52.5} & \num{116} & \num{17} & \num{164} & \num{685} & \num{18} & \num{0} & \num{182} & \num{3} & \num{141} & \num{673} & \num{1} & \num{0} \\
        \num{1.10} & \num{55.0} & \num{167} & \num{24} & \num{243} & \num{542} & \num{24} & \num{0} & \num{267} & \num{6} & \num{207} & \num{519} & \num{1} & \num{0} \\
        \num{1.10} & \num{55.0} & \num{174} & \num{23} & \num{300} & \num{467} & \num{36} & \num{0} & \num{301} & \num{6} & \num{243} & \num{450} & \num{0} & \num{0} \\
        \num{1.15} & \num{47.5} & \num{137} & \num{18} & \num{232} & \num{584} & \num{29} & \num{0} & \num{234} & \num{6} & \num{188} & \num{571} & \num{1} & \num{0} \\
        \num{1.15} & \num{50.0} & \num{203} & \num{26} & \num{326} & \num{424} & \num{21} & \num{0} & \num{335} & \num{1} & \num{256} & \num{408} & \num{0} & \num{0} \\
        \num{1.15} & \num{55.0} & \num{249} & \num{25} & \num{324} & \num{358} & \num{43} & \num{1} & \num{409} & \num{2} & \num{259} & \num{325} & \num{4} & \num{1} \\
        \num{1.15} & \num{55.0} & \num{289} & \num{28} & \num{321} & \num{357} & \num{45} & \num{0} & \num{418} & \num{3} & \num{264} & \num{314} & \num{0} & \num{1} \\
        \num{1.20} & \num{43.0} & \num{134} & \num{14} & \num{166} & \num{669} & \num{17} & \num{0} & \num{215} & \num{5} & \num{120} & \num{651} & \num{9} & \num{0} \\
        \num{1.20} & \num{45.0} & \num{220} & \num{22} & \num{308} & \num{414} & \num{36} & \num{0} & \num{360} & \num{2} & \num{250} & \num{384} & \num{4} & \num{0} \\
        \num{1.20} & \num{50.0} & \num{258} & \num{33} & \num{325} & \num{344} & \num{39} & \num{0} & \num{418} & \num{13} & \num{264} & \num{303} & \num{1} & \num{0} \\
        \num{1.20} & \num{55.0} & \num{304} & \num{44} & \num{316} & \num{296} & \num{40} & \num{0} & \num{495} & \num{8} & \num{245} & \num{251} & \num{1} & \num{0} \\
        \hline
    \end{tabular}
    \tablecomments{The $N^2$ columns denote the number of evolutions where the runner collides with the specified body; $N^2_\mathrm{None}$ indicate evolutions where the runner did not collide with any body during the entire evolution. The $N^\mathrm{F}$ columns denote the number of evolutions where the runner has been accreted by the specified body; $N^\mathrm{F}_\mathrm{None}$ indicate the number of evolutions where the runner survived until the end of the evolution.}
    \label{tab:res-dyn-earth}
    \end{center}
\end{table*}

\begin{table*}
    \begin{center}
    \caption{Outcome of dynamical evolution for collisions with Venus.}
    \begin{tabular}{|cc|rrrrrr|rrrrrr|}
        \hline
        $\frac{v_\mathrm{coll}}{v_\mathrm{esc}}$ & $\theta_\mathrm{coll}$ & $N_\mathrm{None}^2$ & $N_\mathrm{Merc}^2$ & $N_\mathrm{Venus}^2$ & $N_\mathrm{Earth}^2$ & $N_\mathrm{Mars}^2$ & $N_\mathrm{Jup}^2$ & $N_\mathrm{None}^\mathrm{F}$ & $N_\mathrm{Merc}^\mathrm{F}$ & $N_\mathrm{Venus}^\mathrm{F}$ & $N_\mathrm{Earth}^\mathrm{F}$ & $N_\mathrm{Mars}^\mathrm{F}$ & $N_\mathrm{Jup}^\mathrm{F}$ \\
        & [deg] & & & & & & \num{} & & & & & & \\
        \hline
        \hline
        \num{1.10} & \num{55.0} & \num{76} & \num{13} & \num{776} & \num{126} & \num{9} & \num{0} & \num{121} & \num{1} & \num{760} & \num{118} & \num{0} & \num{0} \\
        \num{1.10} & \num{60.0} & \num{91} & \num{18} & \num{702} & \num{172} & \num{17} & \num{0} & \num{162} & \num{1} & \num{672} & \num{164} & \num{1} & \num{0} \\
        \num{1.15} & \num{47.5} & \num{73} & \num{8} & \num{787} & \num{120} & \num{12} & \num{0} & \num{127} & \num{0} & \num{760} & \num{113} & \num{0} & \num{0} \\
        \num{1.15} & \num{50.0} & \num{91} & \num{23} & \num{686} & \num{184} & \num{16} & \num{0} & \num{171} & \num{2} & \num{657} & \num{170} & \num{0} & \num{0} \\
        \num{1.15} & \num{55.0} & \num{132} & \num{16} & \num{565} & \num{276} & \num{11} & \num{0} & \num{235} & \num{5} & \num{503} & \num{255} & \num{2} & \num{0} \\
        \num{1.15} & \num{60.0} & \num{152} & \num{19} & \num{495} & \num{305} & \num{29} & \num{0} & \num{289} & \num{4} & \num{418} & \num{288} & \num{1} & \num{0} \\
        \num{1.20} & \num{42.5} & \num{57} & \num{10} & \num{809} & \num{112} & \num{12} & \num{0} & \num{90} & \num{3} & \num{789} & \num{108} & \num{10} & \num{0} \\
        \num{1.20} & \num{45.0} & \num{103} & \num{20} & \num{664} & \num{199} & \num{14} & \num{0} & \num{188} & \num{2} & \num{625} & \num{182} & \num{3} & \num{0} \\
        \num{1.20} & \num{50.0} & \num{162} & \num{25} & \num{478} & \num{312} & \num{22} & \num{1} & \num{273} & \num{2} & \num{429} & \num{290} & \num{5} & \num{1} \\
        \num{1.20} & \num{55.0} & \num{204} & \num{31} & \num{468} & \num{264} & \num{33} & \num{0} & \num{366} & \num{5} & \num{386} & \num{241} & \num{1} & \num{1} \\
        \hline
    \end{tabular}
    \tablecomments{Columns are the same as in Table~\ref{tab:res-dyn-earth}.}
    \label{tab:res-dyn-venus}
    \end{center}
\end{table*}

\begin{figure}
	\centering
	\includegraphics{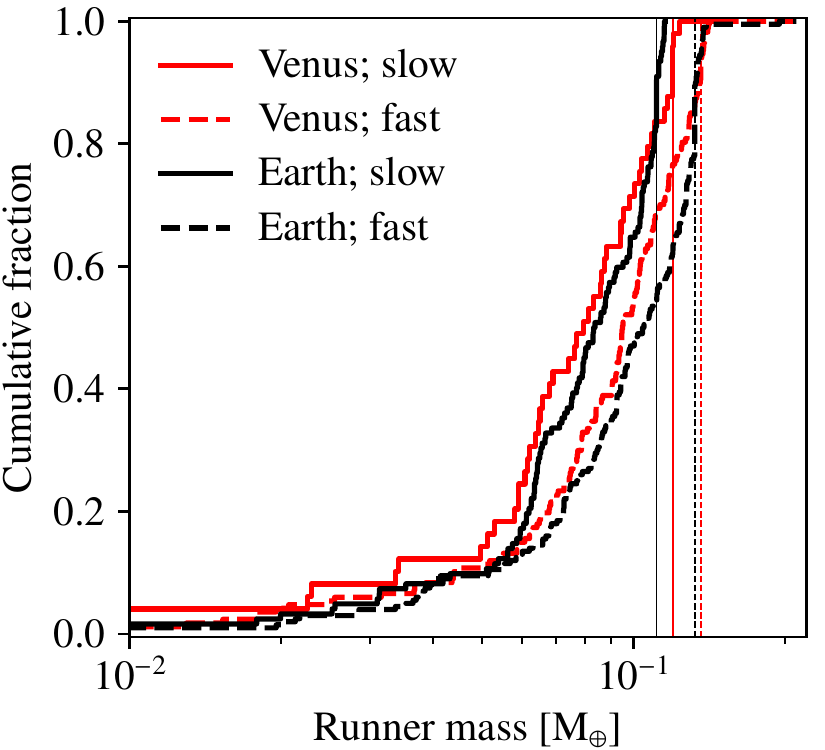}
	\caption{Cumulative distribution of runner masses at the moment they are accreted, for four sets of dynamical evolutions. The mass of the runner after the initial collision for each set is shown with thin vertical lines. Only runners who had at least one intermediate collision are shown, as otherwise they have same mass as that after the initial collision, by definition.}
	\label{fig:runner-acc-mass}
\end{figure}

\begin{figure}
	\centering
	\includegraphics{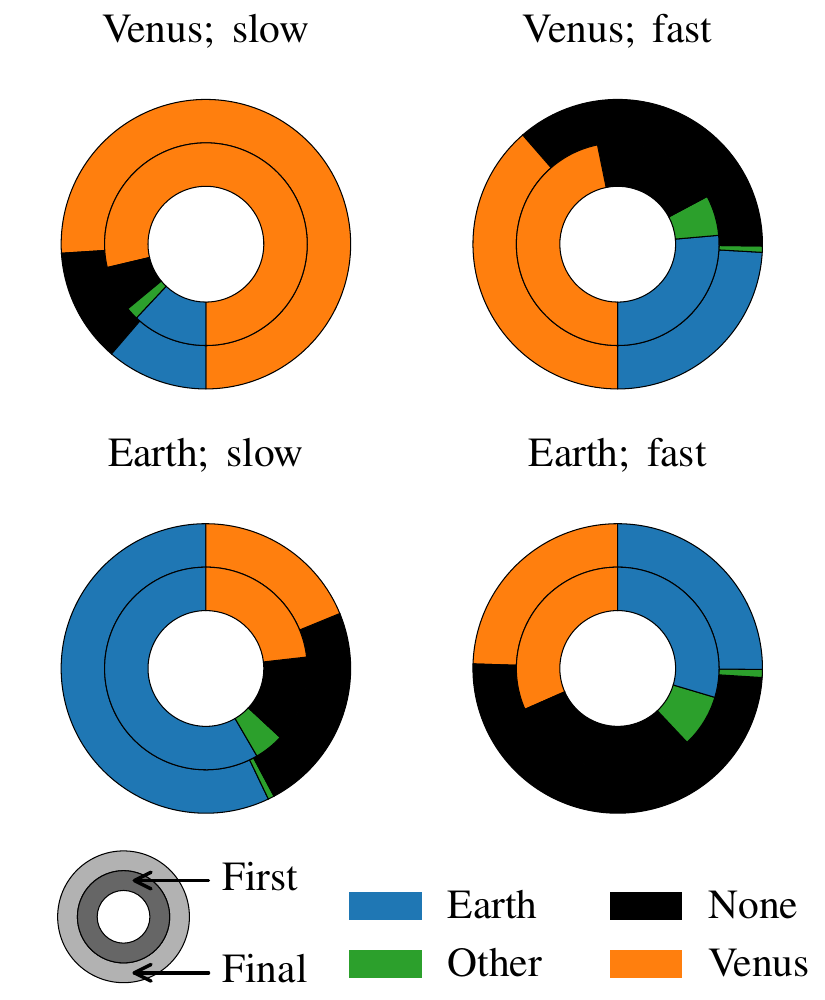}
	\caption{Pie charts of selected results from Tables~\ref{tab:res-dyn-earth} and~\ref{tab:res-dyn-venus}, representing the occurrences for which a body is the target of the first subsequent collision (inner circle) and the body which finally accretes the runner (outer circle). Four series of dynamical evolutions are shown: Venus with $v_\mathrm{coll}/v_\mathrm{esc}=1.15$ and $\theta_\mathrm{coll}=\ang{47.5}$ (upper left), Venus with $v_\mathrm{coll}/v_\mathrm{esc}=1.20$ and $\theta_\mathrm{coll}=\ang{55.0}$ (upper right), Earth with $v_\mathrm{coll}/v_\mathrm{esc}=1.15$ and $\theta_\mathrm{coll}=\ang{47.5}$ (lower left), Earth with $v_\mathrm{coll}/v_\mathrm{esc}=1.20$ and $\theta_\mathrm{coll}=\ang{55.0}$ (lower right). For this sample, Venus accretes nearly all of its slowest runners, and Earth accretes a bit more than half. Venus accretes about half of its faster runners, and a quarter go to Earth. Earth accretes only a quarter of its faster runners and Venus another quarter, for this example.}
	\label{fig:dests}
\end{figure}

\begin{figure*}
	\centering
	\includegraphics{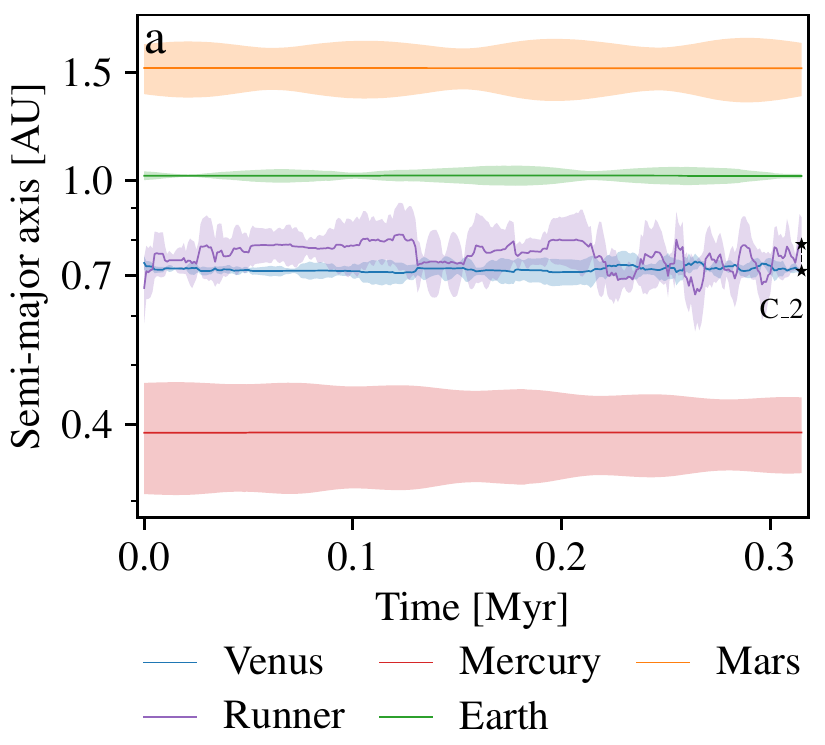}
	\includegraphics{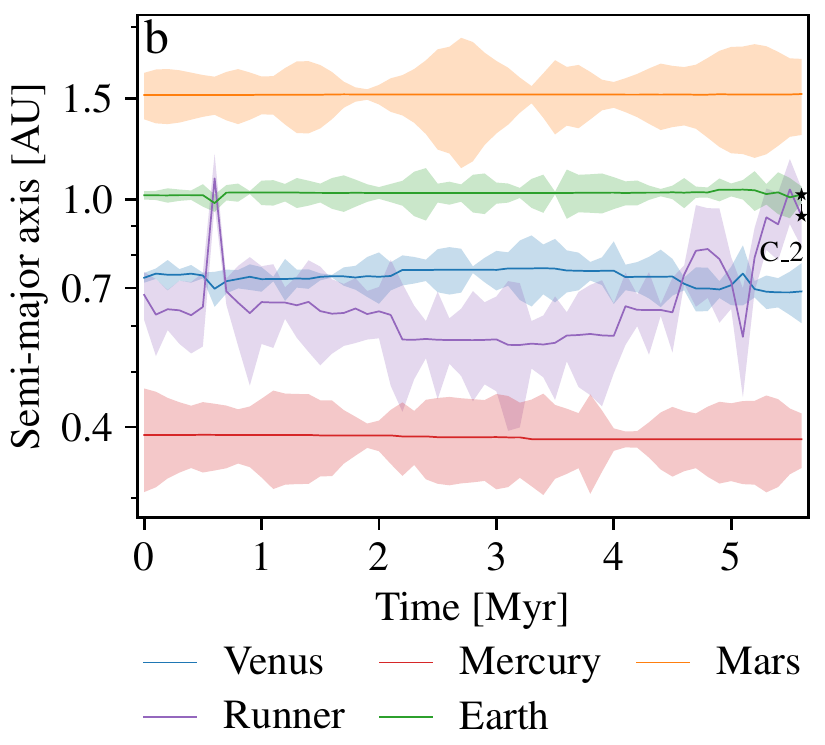}
	\includegraphics{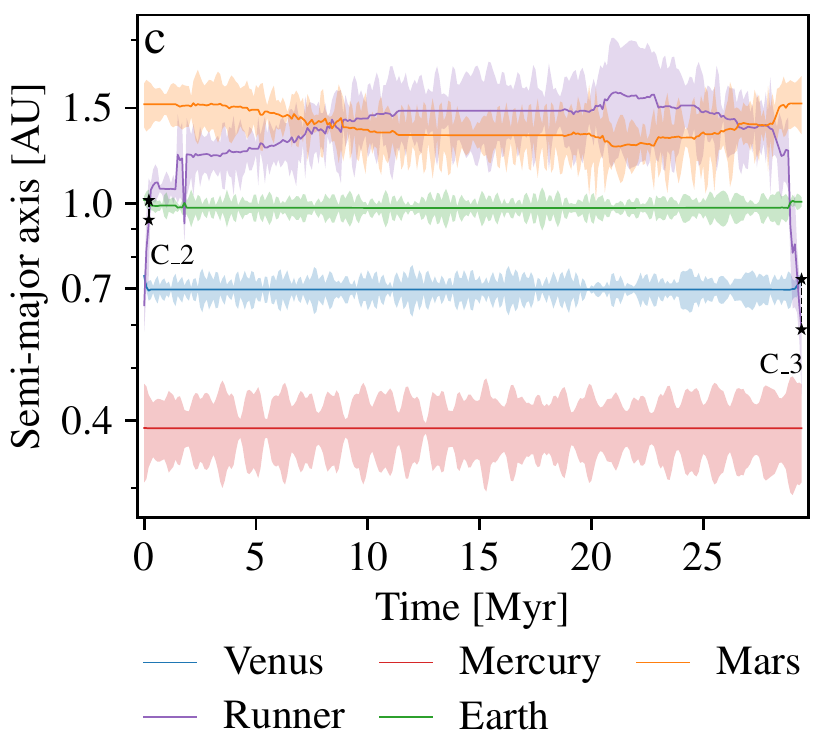}
	\includegraphics{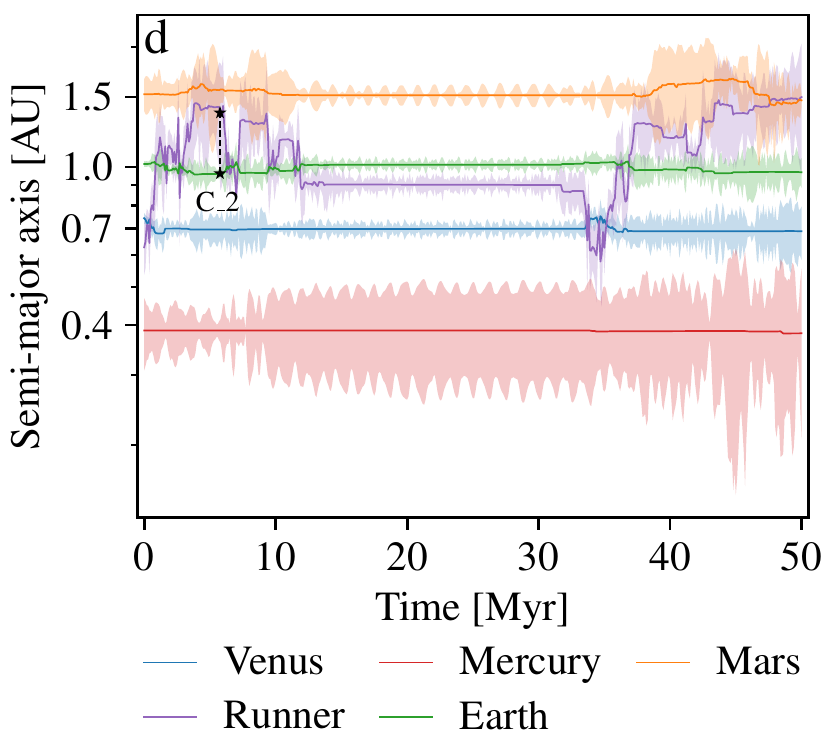}
	\caption{Tracks of selected evolutions following collisions with proto-Venus. The lines show the semi-major axis of the several planets, while the shaded regions denote the perihelion to aphelion. Black stars and the connecting dashed lines show colliding bodies and time, along with the collision identifier. Panel a: runner return on proto-Venus after roughly \SI{3e5}{\year}. Panel b: runner accreted by the proto-Earth after almost \SI{6e6}{\year}. Panel c: hit-and-run collision with the proto-Earth then accreted by proto-Venus after almost \SI{3e7}{\year}. Panel d: hit-and-run with the proto-Earth, survives until \SI{5e7}{\year} (end of \textit{N}-body integration).}
	\label{fig:tracks-venus}
\end{figure*}

Here, we analyze the \textit{N}-body evolutions for the faster-than-$v_{\rm esc}$ remnants of the collisions described in Section \ref{sec:res-initial}. As in \paperone, we have included the other planets on their present orbits, out to Saturn. In these calculations we use the surrogate model of \citet{2019ApJCambioni} to extend the calculation until the runner is lost, either by accretion or by ejection, whereas in \paperone{} the \textit{N}-body evolutions were only integrated up to the first encounter. To compare the demographics of the results of the two methods, we provide in Tables~\ref{tab:res-dyn-earth} and~\ref{tab:res-dyn-venus} the number of evolutions comparing the bodies encountered on the next collision (C\textsubscript{2}; shown with variables $N^2$) versus the bodies onto which the runner was finally accreted (C\textsubscript{F}; shown with variables $N^\mathrm{F}$).

For the hit-and-runs that we modeled, the runners are more massive than Mercury and Mars, so accretion with those planets is actually a case of Mercury or Mars being accreted \emph{by} the runner. In this kind of situation the final result is therefore counted as a survival of the runner (unless it has subsequently been accreted by another more massive planet). In a smaller number of cases the runner is accreted by these planets after it has had an intermediate hit-and-run with another planet, which resulted in mass loss such that the successor of the runner is less massive than Mars or Mercury at the moment of the collision. To show the evolution of the runner's mass by intermediate collisions, we present in Figure~\ref{fig:runner-acc-mass} the cumulative distribution of their masses at the moment of the final collision (C\textsubscript{F}) for the simulations. To focus on the processing by intermediate collisions, only runners having undergone such events are shown. Most runners do not undergo such intermediate collisions (as shown in Figure~\ref{fig:dests}) and their mass is identical to that of the final bodies from the SPH models. The results show that most runners undergoing intermediate collision lose mass, which is consistent with hit-and-run collisions generally resulting in the transfer of some material from the impactor onto the target.

As seen in \paperone, we obtain that the outcomes are principally determined by the relative velocity after the initial collision. To simplify the presentation of the results, we now select four sets of dynamical evolutions that we will discuss in more detail during the remainder of this work, two from each set of collisions with proto-Earth and proto-Venus. For each target planet we select one hit-and-run with a slow runner (with $v_\mathrm{coll}/v_\mathrm{esc}=1.15$ and $\theta_\mathrm{coll}=\ang{47.5}$) and one with a faster runner (with $v_\mathrm{coll}/v_\mathrm{esc}=1.20$ and $\theta_\mathrm{coll}=\ang{55.0}$). The former case was selected because it is the lowest-velocity collision (for the given angle range) where hit-and-run occurs (at lower velocities the runner has an egress velocity less than the mutual escape velocity, resulting in a merger). The latter case is the fastest runner that we modeled. The results of the other sets of dynamical evolution usually lie in between the two that were selected.

A graphical representation of the results is provided in Figure~\ref{fig:dests}, where the inner ring corresponds to the destination of the first collision (C\textsubscript{2}) following the hit-and-run (which is C\textsubscript{1}), and the outer ring is the destination of the final accretion C\textsubscript{F} (the last collision of the chain). Black indicates a runner with no subsequent accretion during the \SI{50}{\mega\year} of our simulations.

The outcome of the first collision is a relatively good predictor of the destination of the final accretion, because most chains end with C\textsubscript{2}. Another common outcome is that the percentage of runners that survive through the end of our simulations (the black section of the outer circle) is larger than the percentage of those that did not undergo any subsequent collision (the black section of the inner circle). This is due to hit-and-runs whose runner survives to the end of evolution, as it happens for those that do not have any collision during the dynamical evolutions.

Comparing similar runners emerging from collisions with proto-Earth and proto-Venus, we find that the ones emerging from proto-Venus are more likely to return and be finally accreted by Venus than vice versa for proto-Earth. Consider our slow runner examples; in these cases about 3/4 of runners are reaccreted by Venus, while \SI{11.3}{\percent} of runners emerging from proto-Venus are accreted by proto-Earth. For the reverse scenario, the percentage of runners from proto-Earth that are accreted by the Earth is significantly smaller, while the probability of ending up at Venus is significantly larger (\SI{18.8}{\percent}). For the faster-runner case, the fraction that are exchanged between the planets increases to \SI{24.5}{\percent} and \SI{24.1}{\percent} respectively, as discussed further below.

We show four systems from the series of evolutions with the ``faster'' runner emerging from proto-Venus ($v_\mathrm{coll}/v_\mathrm{esc}=1.20$ and $\theta_\mathrm{coll}=\ang{55.0}$) in Figure~\ref{fig:tracks-venus}. Each panel represents a common outcome of our evolution, with a runner accretion coming either directly (panel (a)) or after a hit-and-run onto the proto-Earth (panel (c)). Panel (b) shows a case where the runner gets accreted by proto-Earth. Finally, panel (d) shows a runner surviving through the end of the \textit{N}-body evolution, close to the position of Mars after a hit-and-run with the proto-Earth. We see that in most cases of early return on the same planet (akin to panel (a)), the runner remains in the orbital vicinity of the planet between the collisions. However, the runner may wander to a more distant location than what could be implied by only looking at the collisions, with the example of panel (d).

\subsection{Debris production}

\begin{figure}
	\centering
	\includegraphics{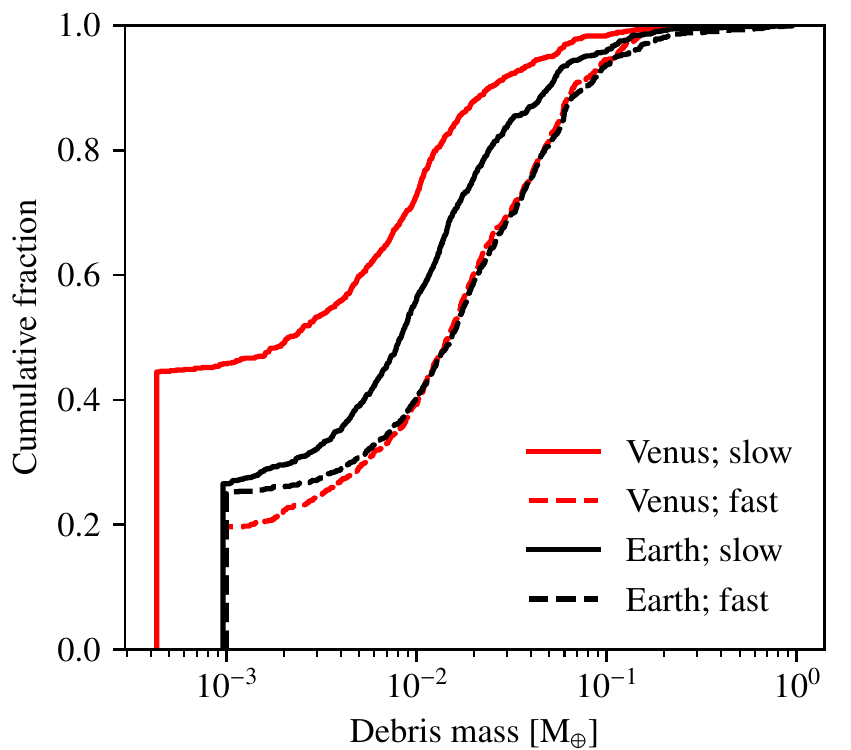}
	\caption{Cumulative distributions of debris produced for four series of dynamical evolutions. These include the debris produced in the initial collision (the $m_\mathrm{lost}$ column in Tables~\ref{tab:res-coll-earth} and~\ref{tab:res-coll-venus}), which is represented as the large jump at the left end of each curve.}
	\label{fig:debris}
\end{figure}

In reality, all collisions generate some level of escaping material, which we do not explicitly track dynamically, and the total amount can be a substantial fraction of the initial planetesimal population mass. To estimate the amount of mass that is lost in this way, we provide in Figure~\ref{fig:debris} the cumulative distributions of the total amount of debris produced for the same four dynamical evolution sets presented in Figure~\ref{fig:dests}. These include the debris produced in the initial collision, which are shown at the left end of each curve.

In all four sets, the mass distributions show that some evolutions do not produce debris at all. For the evolutions with a low relative velocity after the collision (labeled as slow on the figure), this includes cases where the runner is reaccreted without the production of any debris. This happens most for collisions with a velocity of $v_\mathrm{coll}\approx v_\mathrm{esc}$ and impact angle around \ang{45}. In this region of the parameter space, the surrogate collision model indicates that no debris are produced (right panel of Figure 5 in \citealp{2020ApJEmsenhuberA}). For the faster-than-$v_\mathrm{esc}$ evolutions, this is due in part to evolutions that do not produce any subsequent collisions.

For the other cases, the amount of debris produced is generally between \num{e-2} and \SI{e-1}{\mearth}, that is, a few lunar masses. In most cases, the masses are much smaller than the planetary masses; hence, debris reaccretion is generally not a problem here. Only in a few cases do we obtain a total debris mass of \SI{e-1}{\mearth} or more, and this would be greater for more energetic collision chains.

Neglecting the debris in our simulations can affect the simulations in two ways. First, debris affect the orbits of the planets \citep{2019ApJKobayashi}. Small debris can provide dynamical friction, resulting in lower collision velocities. Second, debris reaccretion affects the boundaries of the different collision regimes. This effect depends on how reaccreted debris are distributed between the remnants. Proportionally more accretion on the largest remnant results in more dissimilar bodies, which would reduce the amount of hit-and-run collisions, and vice versa. It should nevertheless be noted that re-impact by debris need not result in net accretion. There is no simple prescription to determine how debris affects the evolutions; taking all their effects into account is a difficult task that needs to be treated carefully.

\subsection{Length of collision chains}
\label{sec:res-length}

\begin{figure*}
	\centering
	\includegraphics{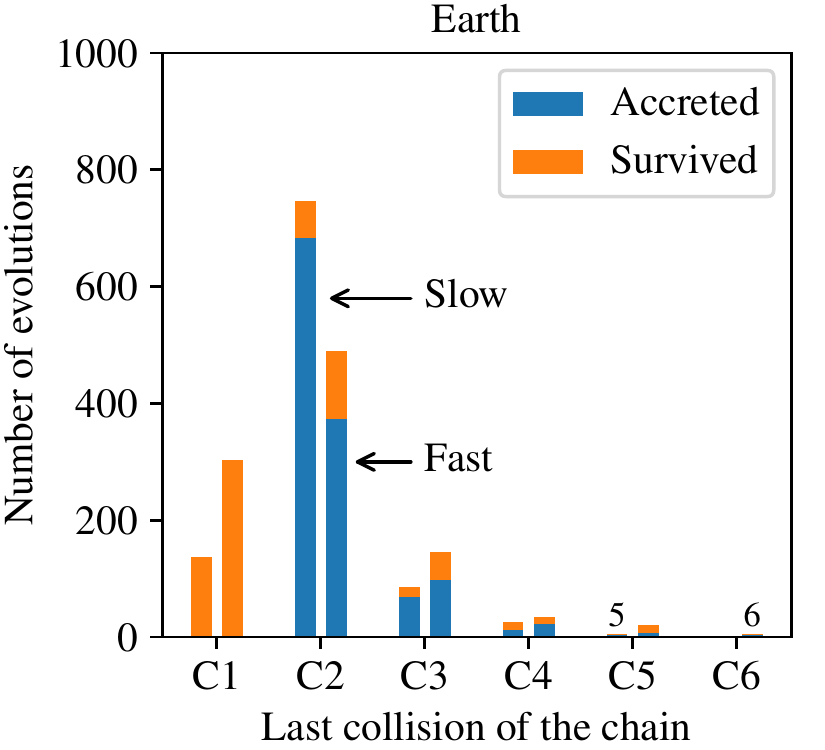}
	\includegraphics{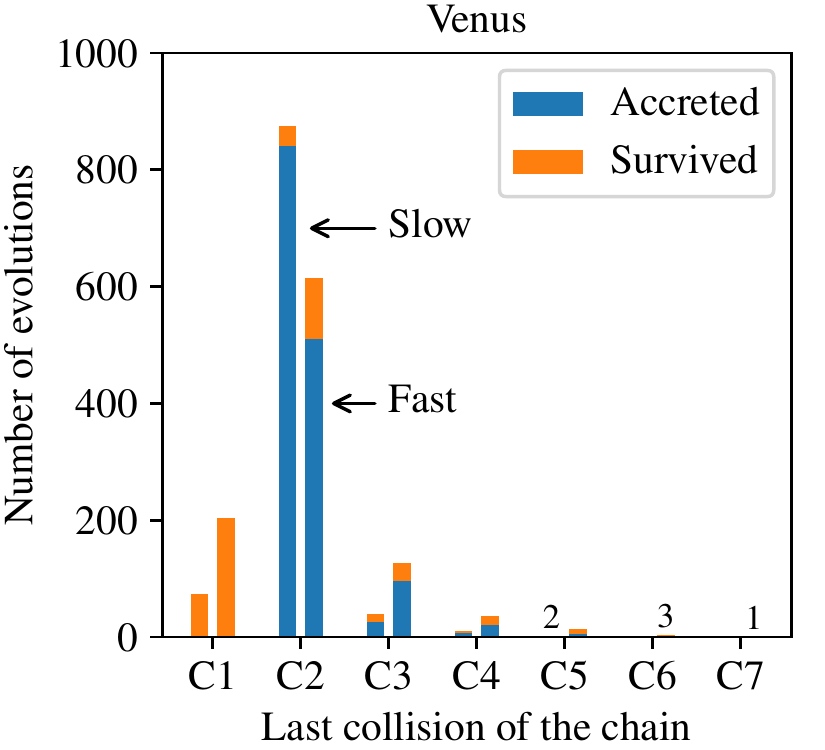}
	\caption{Histograms of the length of collision chains for runners emerging from an hit-and-run (C\textsubscript{1}) with Earth (left) or Venus (right) for two sets of dynamical evolutions each: slow ($v_\mathrm{coll}/v_\mathrm{esc}=1.15$; $\theta_\mathrm{coll}=\ang{47.5}$) and fast ($v_\mathrm{coll}/v_\mathrm{esc}=1.20$; $\theta_\mathrm{coll}=\ang{55.0}$). Initial hit-and-runs that do not result in a return collision are denoted ``C\textsubscript{1}''; Numerical values for C\textsubscript{1} are reported under the column $N_\mathrm{None}^2$ of Tables~\ref{tab:res-dyn-earth} and~\ref{tab:res-dyn-venus}. For each collision, the left and right bar indicates the `slow' and `fast' runner case respectively. Bars that have at least 1 but less than 10 items have the total number shown above them.}
	\label{fig:hist}
\end{figure*}

To understand the number of hit-and-runs that precede an eventual accretion, we provide histograms for the length $N$ of the collision chains in Figure~\ref{fig:hist} for the same four scenarios we presented in Figure~\ref{fig:dests}. In all cases, the most common outcome of these slow hit-and-runs is a follow-up collision that accretes the runner (C\textsubscript{2}). We observe that fast runners tend to have longer chains (more follow-up collisions). The reason for that is linked to our findings from \paperone, that is, the median impact velocity of a follow-up collision is similar to its egress velocity from the previous collision. Thus, fast runners will naturally lead to faster C\textsubscript{2} collisions, which are more likely to result in hit-and-runs. As noted, the preponderance of single-link chains ($N=2$) observed in this analysis accounts for the similar distribution between the initial collision and final accretion observed in Figure~\ref{fig:dests}.

We observe that the fraction of runners that are accreted at each link of the chain decreases. For example, the proportion of surviving runners in the second chain (C\textsubscript2) in Figure~\ref{fig:hist} (left) is small (\SI{21}{\percent} for the slow chain and \SI{46}{\percent} for the faster one). Survival becomes the dominant outcome by the fourth collision in the chain (C\textsubscript4) (\SI{61}{\percent} and \SI{63}{\percent}, respectively). This is, in part, due to the natural bias that for any subsequent collision to result in hit-and-run, it must happen at sufficiently high velocity to overcome the merging velocity threshold.

Another, less straightforward effect arises from the gravitational stirring that occurs between each chain link (as the runner evolves dynamically before re-impacting). As described in \paperone, the median follow-up impact velocity after a hit-and-run is near the egress velocity; however, a tail end of the distribution extends to much higher velocities from \textit{N}-body dynamics. For example, in the low-velocity hit-and-run chains for proto-Venus (Figure~\ref{fig:hist}, right), the median value for impact velocities in C\textsubscript{2} is $v_\mathrm{coll}/v_\mathrm{esc}=1.01$. However, the survivors (hit-and-runs) of C\textsubscript{2} have a median impact velocity of $v_\mathrm{coll}/v_\mathrm{esc}=1.21$, biasing the subsequent collisions to higher velocities.

This bias is compounded at each link of the chain, with the median survivor velocity increasing each time. Longer chains thus increasingly sample less probable, high-velocity scenarios.
The mutual dynamical evolution between chains, which is inherently chaotic, will introduce variability in impact angle and velocity, producing more diverse outcomes in longer chains.

\subsection{Time until loss of the runner}
\label{sec:res-time}

\begin{figure}
	\centering
	\includegraphics{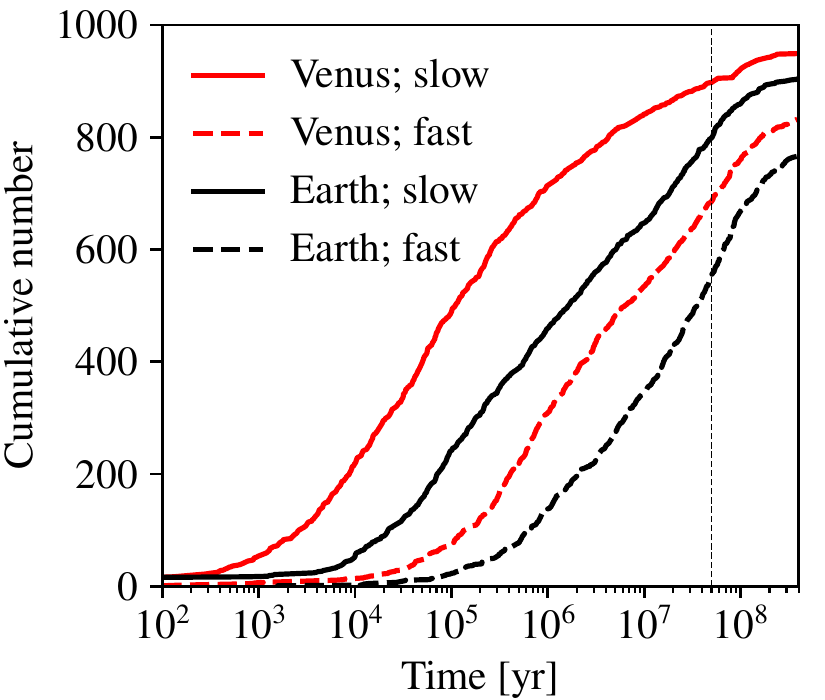}
	\caption{Cumulative distribution across 1,000 simulations of the hit-and-run chain resolution time (the time at which the runner was lost by accretion, either with a planet or the sun, or ejection) for four series of dynamical evolutions: slow runner from proto-Venus (solid red), slow runner from proto-Earth (solid black), fast runner from proto-Venus (dashed red) and fast runner from proto-Earth (dashed black). The vertical dashed line at \SI{50}{\mega\year} is the epoch where we compute the analyses of the chains and collision statistics.}
	\label{fig:time}
\end{figure}

The previous section discussed the differences in the collision chains starting at proto-Earth and proto-Venus, beginning with two sets of hit-and-runs. Earth formation involves somewhat longer chains, for example. It is therefore natural to also compare the timescales for hit-and-run chains to resolve either by accretion or by ejection. In Figure~\ref{fig:time} we show the cumulative resolution time for collision chains. For this figure, the four sets of dynamical evolutions were extended beyond where we computed our collision/chain statistics (at \SI{50}{\mega\year}), out to \SI{400}{\mega\year}. These evolutions are much longer than those in \paperone, as the concern of that work was only about a single follow-up collision.

Through this comparison of the cumulative results of the four sets of simulations, we find that there are distinct differences in the temporal length of hit-and-run chains between the Venus and Earth dynamical zones. At any given epoch, \num{\sim10}--\SI{40}{\percent} more of the hit-and-run chains at \SI{\sim0.7}{\au} (our proto-Venus scenario) tend to have concluded as compared to those at Earth. This is an expected consequence of the shorter collisional timescales of the inner solar system, as well as the lower mass of proto-Venus making merger less probable for a given giant impact. We find that the time required for a collision chain to resolve can be comparable to estimates for the duration of late-stage solar system formation (\SI{\sim200}{\mega\year}; \citealp{2013IcarusChambers,2018MNRASRaymond}). The median time for a chain to conclude in the case of a slow runner at \SI{1}{\au} (Earth) is \SI{1}{\mega\year}. In contrast, the median time for a chain to conclude in the slow-runner case at proto-Venus is only \SI{0.1}{\mega\year}.
These differences are much larger than the ratio of the orbital periods.

Another important factor is that different collisional-dynamical outcomes of a hit-and-run chain dominate at different epochs. Chain resolution at early times is often due to accretion of the runner, whereas at late times chain resolution becomes increasingly dominated by ejection. For instance, there are no ejections or accretions by the Sun before \SI{2}{\mega\year}. We also find that there are no runners that swap from Earth to Venus or Venus to Earth before \SI{e5}{\year}, so very early chain resolution is dominated by accretion on the initial body. Despite this systematic behavior as a function of simulation time, we find that by \SI{50}{\mega\year} the statistics of chain outcomes do not change considerably, justifying our use of \SI{50}{\mega\year} for compiling the probabilities in Tables~\ref{tab:res-dyn-earth} and~\ref{tab:res-dyn-venus} and Figures~\ref{fig:dests} and~\ref{fig:hist}.

\subsection{Exchange of runners}

\begin{figure*}
	\centering
	\includegraphics{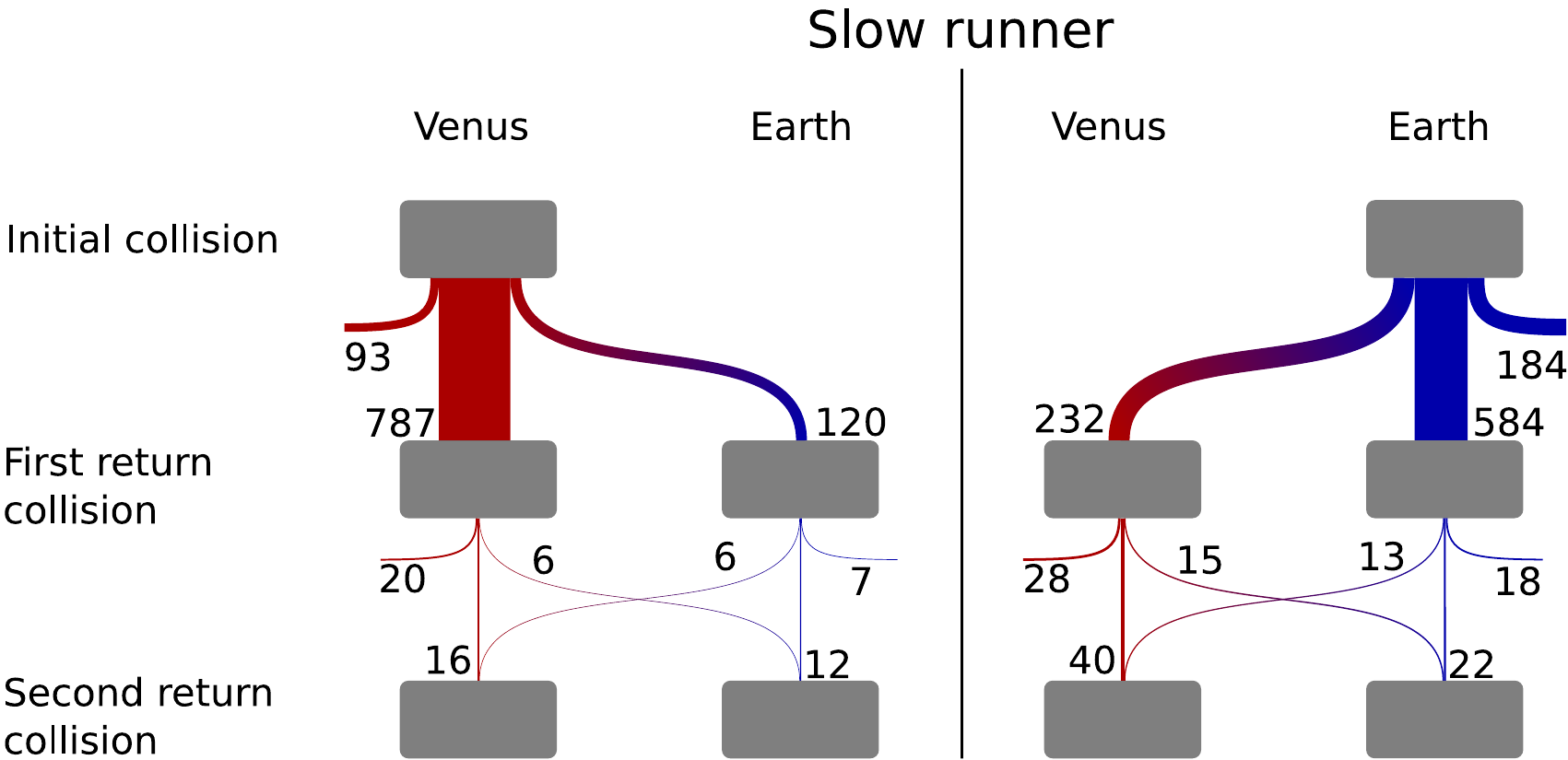}
	\caption{Fate of the ``slow'' runner case, up to the second return collision. The left side is for runners emanating from hit-and-run with proto-Venus while the right side shows runners emanating from hit-and-run with proto-Earth. As an example, among those runners that escape accretion in the initial collisions on Venus, 93 runners are lost, 787 runners re-impact with Venus, and 120 runners impact with Earth. Among those runners that impact with Earth and manage to escape again, 7 runners are lost, 12 runners resolve in mergers at Earth, and 16 runners resolve in mergers at Venus.}
	\label{fig:flow-slow}
\end{figure*}

\begin{figure*}
	\centering
	\includegraphics{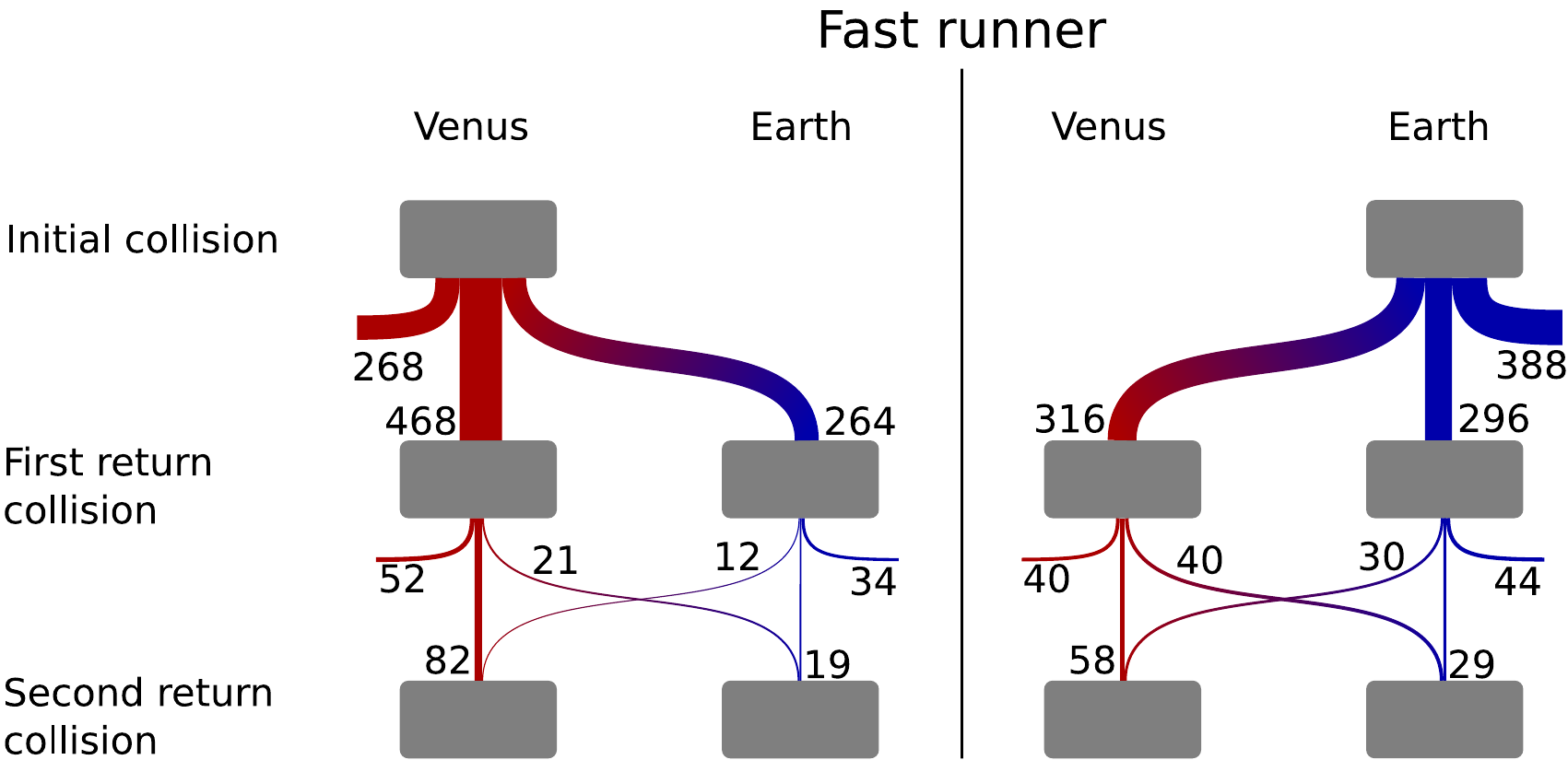}
	\caption{Same as Figure \ref{fig:flow-slow}, but describing the fate of the ``fast''-runner case, up to the second return collision.}
	\label{fig:flow-rapid}
\end{figure*}

In addition to comparing the collisional-dynamical timescales of hit-and-run chains in different dynamical zones, we can also examine the exchange of runners between the two zones. In the pathway diagrams of Figures~\ref{fig:flow-slow} and~\ref{fig:flow-rapid} we show collision chain networks for the same four scenarios examined previously: low impact velocity hit-and-run chains beginning at Venus and Earth (Figure~\ref{fig:flow-slow}, left and right, respectively), and higher impact velocity hit-and-run chains beginning at Venus and Earth (Figure~\ref{fig:flow-rapid}, left and right, respectively). Numbers are per 1000 cases. The patterns that were obtained in the previous sections are reflected here, especially that low impact velocity runners emanating from proto-Venus have a higher rate of return to Venus, with a small fraction transported to Earth, i.e. Venus holds on to its runners. In contrast to this, almost twice as many low impact velocity runners leaving proto-Earth end up colliding with Venus. As expected, the higher-velocity runners show an even greater proportion of transport between the dynamical zones and a greater probability of a second return collision, although, as before, Venus is much more efficient than Earth at retaining its runners.

In all cases, only a small overall fraction of runners ($\lesssim$\SI{10}{\percent}) survive past the first follow-up collision, independently of the target of that collision. Importantly, we also notice that the total fraction of second return collisions in the hit-and-run chains originating at Venus changes considerably with initial impact velocity (\SI{3}{\percent}--\SI{10}{\percent}), whereas hit-and-run chains originating at Earth are less sensitive to the initial impact velocity (increasing from \SI{6}{\percent} to \SI{9}{\percent}). This indicates that further study is needed for hit-and-runs in the velocity range $\sim 1.3-1.6 v_\mathrm{esc}$ and faster.

We note an evolution in the dynamical character of runners that survive subsequent collisions. A larger percentage of the runners that survive C\textsubscript{2} do not collide later with either Earth or Venus, compared to the original runners egressing from C\textsubscript{1}. Out of the 42 slow runners that survive C\textsubscript{2} with proto-Venus (left panel of Figure~\ref{fig:flow-slow}), only 22 of them (\SI{52}{\percent}) were found to have a subsequent collision with proto-Earth or proto-Venus, compared to \SI{91}{\percent} of the runners from C\textsubscript{1}. As we discussed in Section~\ref{sec:res-length}, the median impact velocity of each subsequent collision increases, which results in a greater likelihood of interacting with other planets, decreasing the likelihood of re-impacting the initial target.

For Earth, there is a greater overall transfer of runners compared to Venus and a likelihood of longer chains. We therefore find it more probable that runners from proto-Earth would end up in the inner reaches of the solar system than vice versa. Indeed, as can be seen in Figure~\ref{fig:flow-rapid}, higher-velocity hit-and-runs into proto-Earth are less likely to resolve in mergers at Earth than they are to resolve at Venus, the next planet in. The opposite is true for proto-Venus, which tends to resolve chains at Venus in both the slow and faster scenarios.

In summary, we find that in the case of hit-and-runs with Earth at \SI{1.0}{\au} with a departing runner velocity $v_\mathrm{dep} \gtrsim 1.1 v_\mathrm{esc}$, \emph{the runner is about as likely to collide with Venus as it is to return to Earth}. The converse is not true: runners from hit-and-runs into proto-Venus, at least for the ``slow'' hit-and-runs we have studied, have a significantly greater likelihood of returning to Venus than making it to Earth. Venus holds onto its runners and is a sink for runners whose planetary parent bodies began farther our in the solar system.

\section{Conclusion}

In this work, we aim to understand whether the geophysical differences between Venus and Earth can be potentially explained by a systematic difference in the giant impacts that formed them. To study this, we perform dynamical evolutions of remnants of hit-and-run collisions until the runner is finally accreted or ejected. The runners may experience subsequent hit-and-runs, a \textit{collision chain} C\textsubscript{1}, C\textsubscript{2}, ... The first C\textsubscript{1} is modeled using our updated SPH code, and subsequent collisions in the chain are represented in an \textit{N}-body routine \citep{2012SoftwareChambers} using the surrogate giant impact model described in \citet{2019ApJCambioni} and the methodology of \citet{2020ApJEmsenhuberA}. The starting collisions take place at \SI{1.0}{\au} and \SI{0.7}{\au}, respectively.

We arrive at four main conclusions:
\begin{itemize}
    \item The terrestrial planets are not isolated during the late stage of planetary formation. Runners emerging from hit-and-runs with one planet are likely to collide with another, with varying probability depending several factors, such as relative velocity and orbital configuration.
    \item Long chains are less probable because they require larger initial velocity, and higher-velocity runners are less likely to return.
    \item Earth serves as sort of a vanguard for Venus, capable of slowing late-stage projectiles down in this manner, yet not accreting more than about half of them itself.
    \item Earth runners end up at Venus with the same likelihood that they return to Earth. Venus, however, retains the majority of its runners in all situations we studied, effectively serving as a sink in these scenarios.
\end{itemize}

Moderate-velocity hit-and-run collisions are akin to dissipative interactions, acting to slow down impacting planets. If the egress velocity is just above the mutual escape velocity, hit-and-run collisions can lead to subsequent accretionary collisions, as the collision velocity of C\textsubscript{2} is usually close to the egress velocity from C\textsubscript{1} (\paperone). But they can also act like a headwind for the runners, leading to a reduction in energy and an overall inward shift of their orbits.

So, if the terrestrial planets formed in multiple giant impacts, then \textit{Venus is significantly more likely than the Earth to have accreted a massive outer solar system body during the late stage of planet formation.}
The Earth, by contrast, has no terrestrial planet beyond its orbit to act as a vanguard. Mars is about the same mass as the late-stage projectiles \citep{2017GRLPiet}, \SI{0.1}{\mearth}, and thus relatively inconsequential in terms of slowing them down through hit-and-run, so Earth has to do it on its own.

These demographic differences have broad implications for Earth and Venus formation specifically, and terrestrial planet formation in general, that require further exploration.
The accretion of most runners by Venus, and fewer by Earth, and the handing-off of faster runners from Earth to Venus are a newly identified aspect of their late-stage formation that would influence Venus's bulk composition and could potentially lead to substantial differences in their final spin states, core-mantle dynamics, and satellite formation which is the subject of the next paper in this series \citep{2021PSJAsphaug}.

\begin{acknowledgments}
A.E., E.A., S.C., and S.R.S. acknowledge support from NASA under grant 80NSSC19K0817 and the University of Arizona.
T.S.J.G. acknowledges support from the Arizona State University Space Technology and Science (``NewSpace'') Initiative.
We thank the anonymous reviewers, whose comments helped to improve the manuscript.
An allocation of computer time from the UA Research Computing High Performance Computing (HPC) is gratefully acknowledged.
\end{acknowledgments}

\software{Mercury \citep{1999MNRASChambers}, collscaling \citep{2019ApJCambioni,2020ApJEmsenhuberA}, matplotlib \citep{2007CSEHunter}}

\bibliographystyle{aasjournal}
\bibliography{manu,add}

\begin{thebibliography}{}
\expandafter\ifx\csname natexlab\endcsname\relax\def\natexlab#1{#1}\fi
\providecommand{\url}[1]{\href{#1}{#1}}
\providecommand{\dodoi}[1]{doi:~\href{http://doi.org/#1}{\nolinkurl{#1}}}
\providecommand{\doeprint}[1]{\href{http://ascl.net/#1}{\nolinkurl{http://ascl.net/#1}}}
\providecommand{\doarXiv}[1]{\href{https://arxiv.org/abs/#1}{\nolinkurl{https://arxiv.org/abs/#1}}}

\bibitem[{{Agnor} {et~al.}(1999){Agnor}, {Canup}, \&
  {Levison}}]{1999IcarusAgnor}
{Agnor}, C.~B., {Canup}, R.~M., \& {Levison}, H.~F. 1999, \icarus, 142, 219,
  \dodoi{10.1006/icar.1999.6201}

\bibitem[{{Asphaug}(2010)}]{2010ChEGAsphaug}
{Asphaug}, E. 2010, Chemie der Erde / Geochemistry, 70, 199,
  \dodoi{10.1016/j.chemer.2010.01.004}

\bibitem[{{Asphaug}(2014)}]{2014AREPSAsphaug}
---. 2014, \areps, 42, 551, \dodoi{10.1146/annurev-earth-050212-124057}

\bibitem[{{Asphaug} {et~al.}(2006){Asphaug}, {Agnor}, \&
  {Williams}}]{2006NatureAsphaug}
{Asphaug}, E., {Agnor}, C.~B., \& {Williams}, Q. 2006, \nat, 439, 155,
  \dodoi{10.1038/nature04311}

\bibitem[{{Asphaug} {et~al.}(2015){Asphaug}, {Collins}, \&
  {Jutzi}}]{2015AstIVAsphaug}
{Asphaug}, E., {Collins}, G., \& {Jutzi}, M. 2015, in Asteroids IV, ed.
  P.~{Michel}, F.~E. {DeMeo}, \& W.~F. {Bottke} (University of Arizona Press),
  661--677

\bibitem[{Asphaug {et~al.}(2021)Asphaug, Emsenhuber, Cambioni, \&
  Gabriel}]{2021PSJAsphaug}
Asphaug, E., Emsenhuber, A., Cambioni, S., \& Gabriel, T. S.~J. 2021, \psj, 2,
  200, \dodoi{10.3847/PSJ/ac19b2}

\bibitem[{{Asphaug} \& {Reufer}(2014)}]{2014NatGeoAsphaug}
{Asphaug}, E., \& {Reufer}, A. 2014, \natgeo, 7, 564, \dodoi{10.1038/ngeo2189}

\bibitem[{{Auclair-Desrotour} {et~al.}(2017){Auclair-Desrotour}, {Laskar},
  {Mathis}, \& {Correia}}]{2017AAAuclairDesrotour}
{Auclair-Desrotour}, P., {Laskar}, J., {Mathis}, S., \& {Correia}, A.~C.~M.
  2017, \aap, 603, A108, \dodoi{10.1051/0004-6361/201628701}

\bibitem[{{Balsara}(1991)}]{1991PhDBalsara}
{Balsara}, D.~S. 1991, PhD thesis, Illinois Univ., Urbana-Champaign.

\bibitem[{{Barnes} \& {Hut}(1986)}]{1986NatureBarnesHut}
{Barnes}, J., \& {Hut}, P. 1986, \nat, 324, 446, \dodoi{10.1038/324446a0}

\bibitem[{{Benz}(1991)}]{1991LNPBenz}
{Benz}, W. 1991, in Lecture Notes in Physics, Berlin Springer Verlag, Vol. 373,
  Late Stages of Stellar Evolution. Computational Methods in Astrophysical
  Hydrodynamics, ed. C.~B. {De Loore}, 258

\bibitem[{{Cambioni} {et~al.}(2019){Cambioni}, {Asphaug}, {Emsenhuber},
  {Gabriel}, {Furfaro}, \& {Schwartz}}]{2019ApJCambioni}
{Cambioni}, S., {Asphaug}, E., {Emsenhuber}, A., {et~al.} 2019, \apj, 875, 40,
  \dodoi{10.3847/1538-4357/ab0e8a}

\bibitem[{{Cambioni} {et~al.}(2021){Cambioni}, {Jacobson}, {Emsenhuber},
  {Asphaug}, {Rubie}, {Gabriel}, {Schwartz}, \& {Furfaro}}]{2021PSJCambioni}
{Cambioni}, S., {Jacobson}, S.~A., {Emsenhuber}, A., {et~al.} 2021, The
  Planetary Science Journal, 2, 93, \dodoi{10.3847/PSJ/abf0ad}

\bibitem[{{Campbell} {et~al.}(2019){Campbell}, {Campbell}, {Carter},
  {Chandler}, {Giorgini}, {Margot}, {Morgan}, {Nolan}, {Perillat}, \&
  {Whitten}}]{2019IcarusCampbell}
{Campbell}, B.~A., {Campbell}, D.~B., {Carter}, L.~M., {et~al.} 2019, \icarus,
  332, 19, \dodoi{10.1016/j.icarus.2019.06.019}

\bibitem[{{Canup}(2004)}]{2004IcarusCanup}
{Canup}, R.~M. 2004, \icarus, 168, 433, \dodoi{10.1016/j.icarus.2003.09.028}

\bibitem[{{Canup}(2008)}]{2008IcarusCanup}
---. 2008, \icarus, 196, 518, \dodoi{10.1016/j.icarus.2008.03.011}

\bibitem[{{Canup} \& {Asphaug}(2001)}]{2001NatureCanup}
{Canup}, R.~M., \& {Asphaug}, E. 2001, \nat, 412, 708, \dodoi{10.1038/35089010}

\bibitem[{{Chambers}(1999)}]{1999MNRASChambers}
{Chambers}, J.~E. 1999, \mnras, 304, 793,
  \dodoi{10.1046/j.1365-8711.1999.02379.x}

\bibitem[{{Chambers}(2001)}]{2001IcarusChambers}
---. 2001, \icarus, 152, 205, \dodoi{10.1006/icar.2001.6639}

\bibitem[{{Chambers}(2012)}]{2012SoftwareChambers}
---. 2012, {Mercury: A software package for orbital dynamics}.
\newblock \doeprint{1201.008}

\bibitem[{{Chambers}(2013)}]{2013IcarusChambers}
---. 2013, \icarus, 224, 43, \dodoi{10.1016/j.icarus.2013.02.015}

\bibitem[{{Chau} {et~al.}(2018){Chau}, {Reinhardt}, {Helled}, \&
  {Stadel}}]{2018ApJChau}
{Chau}, A., {Reinhardt}, C., {Helled}, R., \& {Stadel}, J. 2018, \apj, 865, 35,
  \dodoi{10.3847/1538-4357/aad8b0}

\bibitem[{{Ciesla}(2009)}]{2009IcarusCiesla}
{Ciesla}, F.~J. 2009, \icarus, 200, 655, \dodoi{10.1016/j.icarus.2008.12.009}

\bibitem[{{Clement} {et~al.}(2018){Clement}, {Kaib}, {Raymond}, \&
  {Walsh}}]{2018IcarusClement}
{Clement}, M.~S., {Kaib}, N.~A., {Raymond}, S.~N., \& {Walsh}, K.~J. 2018,
  \icarus, 311, 340, \dodoi{10.1016/j.icarus.2018.04.008}

\bibitem[{{Correia} \& {Laskar}(2001)}]{2001NatureCorreiaLaskar}
{Correia}, A. C.~M., \& {Laskar}, J. 2001, \nat, 411, 767,
  \dodoi{10.1038/35081000}

\bibitem[{{{\'C}uk} \& {Stewart}(2012)}]{2012ScienceCuk}
{{\'C}uk}, M., \& {Stewart}, S.~T. 2012, Science, 338, 1047,
  \dodoi{10.1126/science.1225542}

\bibitem[{{Dones} \& {Tremaine}(1993)}]{1993IcarusDonesTremaine}
{Dones}, L., \& {Tremaine}, S. 1993, \icarus, 103, 67,
  \dodoi{10.1006/icar.1993.1059}

\bibitem[{{Emsenhuber} \& {Asphaug}(2019{\natexlab{a}})}]{2019ApJEmsenhuberA}
{Emsenhuber}, A., \& {Asphaug}, E. 2019{\natexlab{a}}, \apj, 875, 95,
  \dodoi{10.3847/1538-4357/ab0c1d}

\bibitem[{{Emsenhuber} \& {Asphaug}(2019{\natexlab{b}})}]{2019ApJEmsenhuberB}
---. 2019{\natexlab{b}}, \apj, 881, 102, \dodoi{10.3847/1538-4357/ab2f8e}

\bibitem[{{Emsenhuber} \& {Cambioni}(2019)}]{2019SoftwareEmsenhuberCambioni}
{Emsenhuber}, A., \& {Cambioni}, S. 2019, {collresolve}, 1.1,  Zenodo,
  \dodoi{10.5281/zenodo.3560892}

\bibitem[{{Emsenhuber} {et~al.}(2020){Emsenhuber}, {Cambioni}, {Asphaug},
  {Gabriel}, {Schwartz}, \& {Furfaro}}]{2020ApJEmsenhuberA}
{Emsenhuber}, A., {Cambioni}, S., {Asphaug}, E., {et~al.} 2020, \apj, 891, 6,
  \dodoi{10.3847/1538-4357/ab6de5}

\bibitem[{{Emsenhuber} {et~al.}(2018){Emsenhuber}, {Jutzi}, \&
  {Benz}}]{2018IcarusEmsenhuber}
{Emsenhuber}, A., {Jutzi}, M., \& {Benz}, W. 2018, \icarus, 301, 247,
  \dodoi{10.1016/j.icarus.2017.09.017}

\bibitem[{{Fischer} \& {Ciesla}(2014)}]{2014EPSLFischerCiesla}
{Fischer}, R.~A., \& {Ciesla}, F.~J. 2014, \epsl, 392, 28,
  \dodoi{10.1016/j.epsl.2014.02.011}

\bibitem[{{Frelikh} {et~al.}(2019){Frelikh}, {Jang}, {Murray-Clay}, \&
  {Petrovich}}]{2019ApJLFrelikh}
{Frelikh}, R., {Jang}, H., {Murray-Clay}, R.~A., \& {Petrovich}, C. 2019,
  \apjl, 884, L47, \dodoi{10.3847/2041-8213/ab4a7b}

\bibitem[{{Gabriel} {et~al.}(2020){Gabriel}, {Jackson}, {Asphaug}, {Reufer},
  {Jutzi}, \& {Benz}}]{2020ApJGabriel}
{Gabriel}, T. S.~J., {Jackson}, A.~P., {Asphaug}, E., {et~al.} 2020, \apj, 892,
  40, \dodoi{10.3847/1538-4357/ab528d}

\bibitem[{{Genda} {et~al.}(2012){Genda}, {Kokubo}, \& {Ida}}]{2012ApJGenda}
{Genda}, H., {Kokubo}, E., \& {Ida}, S. 2012, \apj, 744, 137,
  \dodoi{10.1088/0004-637X/744/2/137}

\bibitem[{{G{\'o}rski} {et~al.}(2005){G{\'o}rski}, {Hivon}, {Banday},
  {Wandelt}, {Hansen}, {Reinecke}, \& {Bartelmann}}]{2005ApJGorski}
{G{\'o}rski}, K.~M., {Hivon}, E., {Banday}, A.~J., {et~al.} 2005, \apj, 622,
  759, \dodoi{10.1086/427976}

\bibitem[{{Hansen}(2009)}]{2009ApJHansen}
{Hansen}, B. M.~S. 2009, \apj, 703, 1131, \dodoi{10.1088/0004-637X/703/1/1131}

\bibitem[{{Hunter}(2007)}]{2007CSEHunter}
{Hunter}, J.~D. 2007, Computing in Science and Engineering, 9, 90,
  \dodoi{10.1109/MCSE.2007.55}

\bibitem[{{Ida} \& {Nakazawa}(1990)}]{1990IcarusIdaNakazawa}
{Ida}, S., \& {Nakazawa}, K. 1990, \icarus, 86, 561,
  \dodoi{10.1016/0019-1035(90)90233-Y}

\bibitem[{{Jackson} {et~al.}(2018){Jackson}, {Gabriel}, \&
  {Asphaug}}]{2018MNRASJackson}
{Jackson}, A.~P., {Gabriel}, T. S.~J., \& {Asphaug}, E.~I. 2018, \mnras, 474,
  2924, \dodoi{10.1093/mnras/stx2901}

\bibitem[{{Jacobson} {et~al.}(2017){Jacobson}, {Rubie}, {Hernlund},
  {Morbidelli}, \& {Nakajima}}]{2017EPSLJacobson}
{Jacobson}, S.~A., {Rubie}, D.~C., {Hernlund}, J., {Morbidelli}, A., \&
  {Nakajima}, M. 2017, \epsl, 474, 375, \dodoi{10.1016/j.epsl.2017.06.023}

\bibitem[{{Johansen} {et~al.}(2021){Johansen}, {Ronnet}, {Bizzarro},
  {Schiller}, {Lambrechts}, {Nordlund}, \& {Lammer}}]{2021SciAdvJohansen}
{Johansen}, A., {Ronnet}, T., {Bizzarro}, M., {et~al.} 2021, arXiv e-prints,
  arXiv:2102.08611.
\newblock \doarXiv{2102.08611}

\bibitem[{{Kaula}(1979)}]{1979JGRKaula}
{Kaula}, W.~M. 1979, \jgr, 84, 999, \dodoi{10.1029/JB084iB03p00999}

\bibitem[{{Kobayashi} {et~al.}(2019){Kobayashi}, {Isoya}, \&
  {Sato}}]{2019ApJKobayashi}
{Kobayashi}, H., {Isoya}, K., \& {Sato}, Y. 2019, \apj, 887, 226,
  \dodoi{10.3847/1538-4357/ab5307}

\bibitem[{{Kokubo} \& {Genda}(2010)}]{2010ApJKokubo}
{Kokubo}, E., \& {Genda}, H. 2010, \apjl, 714, L21,
  \dodoi{10.1088/2041-8205/714/1/L21}

\bibitem[{{Kokubo} \& {Ida}(2000)}]{2000IcarusKokuboA}
{Kokubo}, E., \& {Ida}, S. 2000, \icarus, 143, 15,
  \dodoi{10.1006/icar.1999.6237}

\bibitem[{{Kokubo} \& {Ida}(2002)}]{2002ApJKokubo}
---. 2002, \apj, 581, 666, \dodoi{10.1086/344105}

\bibitem[{{Leinhardt} {et~al.}(2010){Leinhardt}, {Marcus}, \&
  {Stewart}}]{2010ApJLeinhardt}
{Leinhardt}, Z.~M., {Marcus}, R.~A., \& {Stewart}, S.~T. 2010, \apj, 714, 1789,
  \dodoi{10.1088/0004-637X/714/2/1789}

\bibitem[{{Leinhardt} \& {Stewart}(2012)}]{2012ApJLeinhardt}
{Leinhardt}, Z.~M., \& {Stewart}, S.~T. 2012, \apj, 745, 79,
  \dodoi{10.1088/0004-637X/745/1/79}

\bibitem[{{Lissauer} \& {Safronov}(1991)}]{1991IcarusLissauerSafronov}
{Lissauer}, J.~J., \& {Safronov}, V.~S. 1991, \icarus, 93, 288,
  \dodoi{10.1016/0019-1035(91)90213-D}

\bibitem[{{Marinova} {et~al.}(2008){Marinova}, {Aharonson}, \&
  {Asphaug}}]{2008NatureMarinova}
{Marinova}, M.~M., {Aharonson}, O., \& {Asphaug}, E. 2008, \nat, 453, 1216,
  \dodoi{10.1038/nature07070}

\bibitem[{{Melosh}(2007)}]{2007M&PSMelosh}
{Melosh}, H.~J. 2007, Meteoritics and Planetary Science, 42, 2079,
  \dodoi{10.1111/j.1945-5100.2007.tb01009.x}

\bibitem[{{Monaghan}(1992)}]{1992ARA&AMonaghan}
{Monaghan}, J.~J. 1992, \araa, 30, 543,
  \dodoi{10.1146/annurev.aa.30.090192.002551}

\bibitem[{{Monaghan}(1997)}]{1997JCPMonaghan}
---. 1997, \jcoph, 136, 298, \dodoi{10.1006/jcph.1997.5732}

\bibitem[{{Morris} \& {Monaghan}(1997)}]{1997JCPMorrisMonaghan}
{Morris}, J.~P., \& {Monaghan}, J.~J. 1997, \jcoph, 136, 41,
  \dodoi{10.1006/jcph.1997.5690}

\bibitem[{{O'Brien} {et~al.}(2006){O'Brien}, {Morbidelli}, \&
  {Levison}}]{2006IcarusOBrien}
{O'Brien}, D.~P., {Morbidelli}, A., \& {Levison}, H.~F. 2006, \icarus, 184, 39,
  \dodoi{10.1016/j.icarus.2006.04.005}

\bibitem[{{Owen}(2004)}]{2004JCPOwen}
{Owen}, J.~M. 2004, \jcoph, 201, 601, \dodoi{10.1016/j.jcp.2004.06.011}

\bibitem[{{Piet} {et~al.}(2017){Piet}, {Badro}, \& {Gillet}}]{2017GRLPiet}
{Piet}, H., {Badro}, J., \& {Gillet}, P. 2017, \grl, 44, 11,770,
  \dodoi{10.1002/2017GL075225}

\bibitem[{{Quintana} {et~al.}(2016){Quintana}, {Barclay}, {Borucki}, {Rowe}, \&
  {Chambers}}]{2016ApJQuintana}
{Quintana}, E.~V., {Barclay}, T., {Borucki}, W.~J., {Rowe}, J.~F., \&
  {Chambers}, J.~E. 2016, \apj, 821, 126, \dodoi{10.3847/0004-637X/821/2/126}

\bibitem[{{Raymond} {et~al.}(2018){Raymond}, {Boulet}, {Izidoro}, {Esteves}, \&
  {Bitsch}}]{2018MNRASRaymond}
{Raymond}, S.~N., {Boulet}, T., {Izidoro}, A., {Esteves}, L., \& {Bitsch}, B.
  2018, \mnras, 479, L81, \dodoi{10.1093/mnrasl/sly100}

\bibitem[{{Raymond} {et~al.}(2009){Raymond}, {O'Brien}, {Morbidelli}, \&
  {Kaib}}]{2009RaymondIcarus}
{Raymond}, S.~N., {O'Brien}, D.~P., {Morbidelli}, A., \& {Kaib}, N.~A. 2009,
  \icarus, 203, 644, \dodoi{10.1016/j.icarus.2009.05.016}

\bibitem[{{Raymond} {et~al.}(2006){Raymond}, {Quinn}, \&
  {Lunine}}]{2006IcarusRaymond}
{Raymond}, S.~N., {Quinn}, T., \& {Lunine}, J.~I. 2006, \icarus, 183, 265,
  \dodoi{10.1016/j.icarus.2006.03.011}

\bibitem[{{Reinhardt} {et~al.}(2020){Reinhardt}, {Chau}, {Stadel}, \&
  {Helled}}]{2020MNRASReinhardt}
{Reinhardt}, C., {Chau}, A., {Stadel}, J., \& {Helled}, R. 2020, \mnras, 492,
  5336, \dodoi{10.1093/mnras/stz3271}

\bibitem[{{Reinhardt} \& {Stadel}(2017)}]{2017MNRASReinhardtStadel}
{Reinhardt}, C., \& {Stadel}, J. 2017, \mnras, 467, 4252,
  \dodoi{10.1093/mnras/stx322}

\bibitem[{{Reufer}(2011)}]{2011PhDReufer}
{Reufer}, A. 2011, PhD thesis, University of Bern

\bibitem[{{Reufer} {et~al.}(2012){Reufer}, {Meier}, {Benz}, \&
  {Wieler}}]{2012IcarusReufer}
{Reufer}, A., {Meier}, M. M.~M., {Benz}, W., \& {Wieler}, R. 2012, \icarus,
  221, 296, \dodoi{10.1016/j.icarus.2012.07.021}

\bibitem[{{Ribeiro} {et~al.}(2020){Ribeiro}, {Morbidelli}, {Raymond},
  {Izidoro}, {Gomes}, \& {Vieira Neto}}]{2020IcarusRibeiro}
{Ribeiro}, R. d.~S., {Morbidelli}, A., {Raymond}, S.~N., {et~al.} 2020,
  \icarus, 339, 113605, \dodoi{10.1016/j.icarus.2019.113605}

\bibitem[{{Rosswog}(2009)}]{2009NARRosswog}
{Rosswog}, S. 2009, \nar, 53, 78, \dodoi{10.1016/j.newar.2009.08.007}

\bibitem[{{Rubie} {et~al.}(2015){Rubie}, {Jacobson}, {Morbidelli}, {O'Brien},
  {Young}, {de Vries}, {Nimmo}, {Palme}, \& {Frost}}]{2015IcarusRubie}
{Rubie}, D.~C., {Jacobson}, S.~A., {Morbidelli}, A., {et~al.} 2015, \icarus,
  248, 89, \dodoi{10.1016/j.icarus.2014.10.015}

\bibitem[{{Safronov}(1969)}]{1969BookSafronov}
{Safronov}, V.~S. 1969, {Evolution of the Protoplanetary Cloud and Formation of
  the Earth and the Planets} (Moscow: Nauka)

\bibitem[{{Sheppard} \& {Trujillo}(2009)}]{2009IcarusSheppard}
{Sheppard}, S.~S., \& {Trujillo}, C.~A. 2009, \icarus, 202, 12,
  \dodoi{10.1016/j.icarus.2009.02.008}

\bibitem[{{Slattery} {et~al.}(1992){Slattery}, {Benz}, \&
  {Cameron}}]{1992IcarusSlattery}
{Slattery}, W.~L., {Benz}, W., \& {Cameron}, A.~G.~W. 1992, \icarus, 99, 167,
  \dodoi{10.1016/0019-1035(92)90180-F}

\bibitem[{{Stewart} \& {Leinhardt}(2012)}]{2012ApJStewart}
{Stewart}, S.~T., \& {Leinhardt}, Z.~M. 2012, \apj, 751, 32,
  \dodoi{10.1088/0004-637X/751/1/32}

\bibitem[{{Thompson} {et~al.}(2019){Thompson}, {Weinberger}, {Keller},
  {Arnold}, \& {Stark}}]{2019ApJThompson}
{Thompson}, M.~A., {Weinberger}, A.~J., {Keller}, L.~D., {Arnold}, J.~A., \&
  {Stark}, C.~C. 2019, \apj, 875, 45, \dodoi{10.3847/1538-4357/ab0d7f}

\bibitem[{Thompson \& Lauson(1972)}]{ANEOS}
Thompson, S.~L., \& Lauson, H.~S. 1972, {Improvements in the CHART-D
  Radiation-hydrodynamic code III: Revised analytic equations of state}, Tech.
  Rep. SC-RR-71 0714, Sandia National Laboratories

\bibitem[{{Timpe} {et~al.}(2020){Timpe}, {Han Veiga}, {Knabenhans}, {Stadel},
  \& {Marelli}}]{2020CACTimpe}
{Timpe}, M.~L., {Han Veiga}, M., {Knabenhans}, M., {Stadel}, J., \& {Marelli},
  S. 2020, Computational Astrophysics and Cosmology, 7, 2,
  \dodoi{10.1186/s40668-020-00034-6}

\bibitem[{{Tonks} \& {Melosh}(1993)}]{1993JGRTonksMelosh}
{Tonks}, W.~B., \& {Melosh}, H.~J. 1993, \jgr, 98, 5319,
  \dodoi{10.1029/92JE02726}

\bibitem[{{Wetherill}(1985)}]{1985ScienceWetherill}
{Wetherill}, G.~W. 1985, \sci, 228, 877, \dodoi{10.1126/science.228.4701.877}

\bibitem[{{Wilhelms} \& {Squyres}(1984)}]{1984NatureWilhelmsSquyres}
{Wilhelms}, D.~E., \& {Squyres}, S.~W. 1984, \nat, 309, 138,
  \dodoi{10.1038/309138a0}

\bibitem[{{Woo} {et~al.}(2021){Woo}, {Stadel}, {Grimm}, \&
  {Brasser}}]{2021ApJWoo}
{Woo}, J. M.~Y., {Stadel}, J., {Grimm}, S., \& {Brasser}, R. 2021, \apjl, 910,
  L16, \dodoi{10.3847/2041-8213/abed56}

\end{thebibliography}

\end{document}